\pdfoutput=1
\documentclass[11pt,a4paper]{article}
\usepackage{jheppub}

\usepackage{graphicx}
\usepackage{epsfig}
\usepackage{color}
\usepackage{comment}
\usepackage{amsmath}
\usepackage{amssymb}
\usepackage{siunitx}
\usepackage{booktabs}
\usepackage{hyperref}
\usepackage{feynmp-auto}
\newcommand{\be}{\begin{equation}}
\newcommand{\ee}{\end{equation}}
\newcommand{\bea}{\begin{eqnarray}}
\newcommand{\eea}{\end{eqnarray}}

\newcommand{\gapp}{\mathrel{\raise.3ex\hbox{$>$}\mkern-14mu
\lower0.6ex\hbox{$\sim$}}}
\newcommand{\lapp}{\mathrel{\raise.3ex\hbox{$<$}\mkern-14mu
\lower0.6ex\hbox{$\sim$}}}
\def\bbox{{\,\lower0.9pt\vbox{\hrule \hbox{\vrule height 0.2 cm
\hskip 0.2 cm \vrule  height 0.2 cm}\hrule}\,}}

\title{Signals of Doomsday III: Cosmological signatures of the late time $U(1)_{EM}$ symmetry breaking}

\author[a,b]{Amartya Sengupta,}
\author[a]{Dejan Stojkovic}

\affiliation[a]{HEPCOS, Department\, of \,Physics, SUNY\, at\, Buffalo, Buffalo, NY\, 14260-1500, USA}
\affiliation[b]{Department of Physics, University of Cincinnati, Cincinnati, Ohio 45221, USA}

\emailAdd{amartyas@buffalo.edu}
\emailAdd{ds77@buffalo.edu}

\abstract{
Of the universe’s original gauge symmetries, only $SU(3)_c$ (quantum chromodynamics) and $U(1)_{\rm EM}$ (electromagnetism) remain unbroken today. There is, however, no reason to assume that these symmetries are permanent. This paper explores the potential astrophysical observational signatures of a late-time breaking of $U(1)_{\rm EM}$. We present a model with a new massive scalar field whose potential supports a first-order phase transition through the nucleation of true-vacuum bubbles. If the propagation of the bubble walls slows down due to interactions with the surrounding matter and radiation, these signals can reach us before the bubble wall itself arrives. Using the vacuum-mismatch method, we calculate the spectrum of particles produced by such a bubble until the terminal velocity is reached. In addition, we show that frictional dissipation at terminal wall velocity generates a large population of thermally produced scalars and massive photons, which continues even after the mismatch channel shuts off. We then use event generators to simulate the decays of the new scalar and the massive photon into Standard Model particles and obtain, as the final result, the energy spectra of photons and neutrinos. Since the dominant final decay products after hadronization and the decay of unstable particles are photons and neutrinos, they act as long-range signatures of the transition. We also estimate the possible lead time of these photon and neutrino signals relative to the arrival of the bubble wall itself, showing that even a modest subluminal wall velocity can in principle provide an observable precursor. For the conservative set of parameters used here, the thermal channel produces a macroscopically large burst of high-energy photons and neutrinos, which could in principle be detectable from sufficiently nearby bubbles with present or future multi-messenger facilities.
}

\begin{document}
\maketitle
\tableofcontents

\section{Introduction}
It is our current understanding that the universe had a sequence of phase transitions in which an initial gauge symmetry group broke — either completely or into a smaller subgroup. These transitions were highly disruptive, fundamentally reshaping the structure of the universe and producing a new phase drastically different from the previous one. Today, only two gauge symmetries remain unbroken: $SU(3)_c$ and $U(1)_{\rm EM}$. However, unless we are some special observers living at the very end of the cosmological sequence of symmetry breaking, there is no guarantee that these symmetries will persist indefinitely. If they break, the universe will undergo another drastic rearrangement, with profound consequences for all life in the universe, including our own.

Previous studies have explored observational signatures of false Higgs vacuum decay \cite{Greenwood:2008qp,Sengupta:2025jah,Coleman:1977py,Coleman:1980aw,Vachaspati:2003vk,Kobzarev:1974cp,Frampton:1976kf,Callan:1977pt,Linde:1977mm,Linde:1979ny,Linde:1980tt,Linde:1981zj,Krive:1976sg,Lee:1974ma,Chang:1975ir,Strumia:2023awj,Canko:2017ebb,PhysRevLett.111.241801,Bentivegna:2017qry,Branchina:2015nda,Branchina:2016bws,Branchina:2018xdh,Branchina:2019tyy,Branchina:2025jou,Burda:2015isa,Burda:2015yfa,Burda:2016mou,Appelquist:1974tg,Dai:2019eei,Dai:2021boq,Degrassi:2012ry,Gregory:2013hja,Alonso:2023jsi,Kawana:2022lba,Kallosh:2003mt,Krauss:2007rx,Kibble:1980mv,Espinosa:2010hh,Espinosa:2016nld,Espinosa:2025ejf,Frieman:1991tu} and the implications of late-time $SU(3)_c$ symmetry breaking \cite{Stojkovic:2007dw,Sengupta:2025cdm,PhysRevD.19.1906,Schafer:1983kc,Slansky:1981tg,Kusenko:1996jn,Bai:2017zhj}. This paper focuses on the astrophysical signatures of late-time $U(1)_{\rm EM}$ symmetry breaking. Direct discussions of spontaneous \(U(1)_{\rm EM}\) breaking are relatively sparse in the literature; a particularly relevant example is Ref.~\cite{West:2017asb}. Since $U(1)_{\rm EM}$ ensures the masslessness of photons, the nature of the universe after such a transition is difficult to predict, but it would almost certainly be inhospitable to life as we know it. Although the probability of this event may be extremely low, the severity of its consequences justifies a thorough investigation into the observable effects of such a phase transition.

We begin by constructing a model for $U(1)_{\rm EM}$ symmetry breaking that remains consistent with current observations. To evade astrophysical and collider constraints, we assume a first-order phase transition, which requires a new massive scalar field to drive the symmetry breaking. We currently live in a false vacuum where $U(1)_{\rm EM}$ is still unbroken and photons are still massless. The first-order phase transition proceeds through the nucleation of true-vacuum bubbles. As the bubble expands, a continuously changing vacuum state creates a mismatch that generates abundant particle production. Our focus is on long-range observational signatures detectable on Earth. We first compute the decays of the massive scalar and the now-massive photons. Since these particles can produce quarks, we use \texttt{Pythia}\cite{Sjostrand:2014zea,Bierlich:2022pfr} to simulate hadronization and subsequent decays into (massless) photons and neutrinos outside of the bubble.

We also show that, in addition to the direct vacuum-mismatch contribution, frictional dissipation of the bubble-wall energy into the surrounding shocked medium can generate a thermal population of the heavy states present in the broken $U(1)_{\rm EM}$ phase. In the benchmark considered here, this thermal component can exceed the direct mismatch contribution by many orders of magnitude and therefore dominate the final observable signal.

Friction from the surrounding gas and matter can slow the expanding bubble wall. We model this interaction by deriving the wall's dynamics in a viscous medium, including its time-dependent proper acceleration which drives particle production. If the propagation of the bubble wall is slowed down due to interaction with surrounding matter and radiation, we may be able to detect these photons and neutrinos before the bubble wall itself arrives. Once the wall reaches terminal velocity and the proper acceleration drops to zero, the direct acceleration-driven vacuum-mismatch emission of detectable photons and neutrinos also stops. Thus, we also show our resulting spectra produced up to that critical point. At the same time, we show that the very same friction responsible for slowing the wall deposits a substantial amount of energy into the surrounding medium, leading to thermal particle production behind the wall. Unlike the direct vacuum-mismatch contribution, this thermal component does not disappear simply because the proper acceleration becomes small, and therefore provides an additional — and potentially dominant — source of observable high-energy photons and neutrinos.

The rest of the paper is organized as follows. We first introduce the $U(1)_{\rm EM}$ symmetry-breaking model and analyze its vacuum structure, bubble nucleation, and the resulting scalar and photon masses. We then discuss particle production from vacuum mismatch, followed by the bubble-wall dynamics in a viscous medium and the associated thermal particle production from frictional dissipation. After that, we study the phenomenology and decay channels of the heavy broken-phase states and use these results to obtain the final photon and neutrino spectra. We also estimate the possible lead time of these signals relative to the arrival of the bubble wall. Finally, we summarize our results and present additional technical details in the appendices.

\section{Model for \texorpdfstring{$U(1)_{EM}$}{U(1){EM}} gauge symmetry breaking }
\label{Model}
We start with a gauge-invariant Lagrangian containing a complex scalar $\Phi_{EM}$ and the $U(1)_{\rm EM}$ gauge field $A_\mu$:
{\small
\begin{equation} \label{L}
\mathcal{L}_{\text{total}} \;=\; -\frac{1}{4}F_{\mu\nu}F^{\mu\nu}
\;+\; (D_\mu\Phi_{EM})^\dagger(D^\mu\Phi_{EM})
\;-\; V (\Phi_{EM})\,,
\end{equation}}
with
\begin{equation}
F_{\mu\nu}=\partial_\mu A_\nu-\partial_\nu A_\mu,\qquad
D_\mu\Phi_{EM}=(\partial_\mu+i\,q\,A_\mu)\,\Phi_{EM}\,,
\end{equation}
and
\begin{equation}\label{Vpotential}
V (\Phi_{EM}) \;=\; +\,m^2\,|\Phi_{EM}|^2\;+\;\lambda\,|\Phi_{EM}|^4\;+\;\frac{\delta}{{\Lambda}^2}\,|\Phi_{EM}|^6\,,\quad
m^2>0,\ \lambda<0,\ \delta>0,
\end{equation}
which admits a first-order phase transition. Here, $m$, $\Lambda$, $\lambda$ and $\delta$ are constants, while $q$ is the magnitude of the gauge charge.

After spontaneous symmetry breaking we parameterize
\begin{equation}
\Phi_{EM}(x)\;=\;\frac{v+h(x)}{\sqrt{2}}\;e^{i\,\theta(x)/v}\,,
\end{equation}
so the covariant kinetic term contains
\begin{equation}
(D_\mu\Phi_{EM})^\dagger(D^\mu\Phi_{EM})
\;\supset\;
\frac{1}{2}\,(\partial_\mu\theta - q\,v\,A_\mu)^2
\;+\;\frac{1}{2}(\partial_\mu h)^2\,.
\end{equation}
In unitary gauge ($\theta\to 0$) this yields the Proca mass term
\begin{equation}
\frac{1}{2}\,q^2 v^2\,A_\mu A^\mu\,,
\end{equation}
so the photon mass in the broken phase is
\begin{equation}
m_\gamma \;=\; q\,v\,.
\end{equation}
The Lagrangian itself remains gauge invariant, while the vacuum expectation value spontaneously breaks the symmetry\cite{Buchmuller:1985jz,Grzadkowski:2010es}. In this process the would-be Goldstone mode $\theta$ is absorbed into the photon field, providing the longitudinal polarization of the resulting massive vector boson.

\subsection{Mass of the scalar field}
\label{sec:scalar-mass}
We now compute the physical mass of the scalar field. For the algebra in the scalar potential it is convenient to trade the complex field modulus for a canonically normalized real variable
\(\phi \equiv |\Phi_{EM}|\).
In unitary gauge the phase is removed, and we expand about the minimum as
\begin{equation}
\phi(x)\;=\;v+\frac{\chi(x)}{\sqrt{2}}\,,
\end{equation}
so that the potential in Eq.~(\ref{Vpotential}) reads
\begin{equation}
V(\phi)\;=\;+\,m^2\,\phi^{2}+\lambda\,\phi^{4}+\frac{\delta}{\Lambda^{2}}\,\phi^{6}\,,\quad m^2>0,\ \lambda<0,\ \delta>0.
\end{equation}
The stationarity condition at the minimum,
\(\left.\frac{dV}{d\chi}\right|_{\chi=0}=0\), gives
\begin{equation}
\frac{1}{\sqrt{2}}\Bigl[\,2m^{2}v+4\lambda v^{3}+\frac{6\delta}{\Lambda^{2}}v^{5}\Bigr]=0
\qquad\Longrightarrow\qquad
m^{2}\;=\;-\,2\lambda\,v^{2}\;-\;\frac{3\delta}{\Lambda^{2}}\,v^{4}\,.
\end{equation}
The physical (canonically normalized) scalar mass follows from the second derivative,
\begin{align}
\left.\frac{d^{2}V}{d\chi^{2}}\right|_{\chi=0}
&= \frac{1}{2}\Bigl[\,2m^{2}+12\lambda v^{2}+\frac{30\delta}{\Lambda^{2}}v^{4}\Bigr] \nonumber\\[2pt]
&= m^{2}+6\lambda v^{2}+\frac{15\delta}{\Lambda^{2}}v^{4}
\;=\;4\lambda\,v^{2}+\frac{12\delta}{\Lambda^{2}}\,v^{4}\,,
\end{align}
where in the last step we used the stationarity relation.

\section{Metastable vacuum structure and bubble nucleation}

We now turn back to the potential for the scalar field, $\Phi_{EM}$,
\begin{equation}
V(\Phi_{EM})=+\,m^2|\Phi_{EM}|^2+\lambda|\Phi_{EM}|^4+\frac{\delta}{\Lambda^2}|\Phi_{EM}|^6,
\quad m^2>0,\ \lambda<0,\ \delta>0.
\end{equation}
We choose the benchmark parameters
\begin{equation}
m^2=6.667\times10^6\;\mathrm{GeV}^2,\quad
\lambda=-3.227,\quad
\delta=1.867,\quad
\Lambda=2304\;\mathrm{GeV},
\end{equation}
which support a standard ``double-well'' potential, and also evade the existing collider limits on the massive scalar fields in the false vacuum.\footnote{We emphasize that the present \(U(1)_{\rm EM}\) benchmark does not assume large Higgs-like mixing of the neutral radial scalar with the Standard Model Higgs sector. Accordingly, collider constraints should not be interpreted as those of a full-strength heavy Higgs boson, but rather as model-dependent bounds on the scalar's production rate and visible branching fractions.}
\begin{figure}[!htbp]
   \centering
\includegraphics[width=10.5cm]{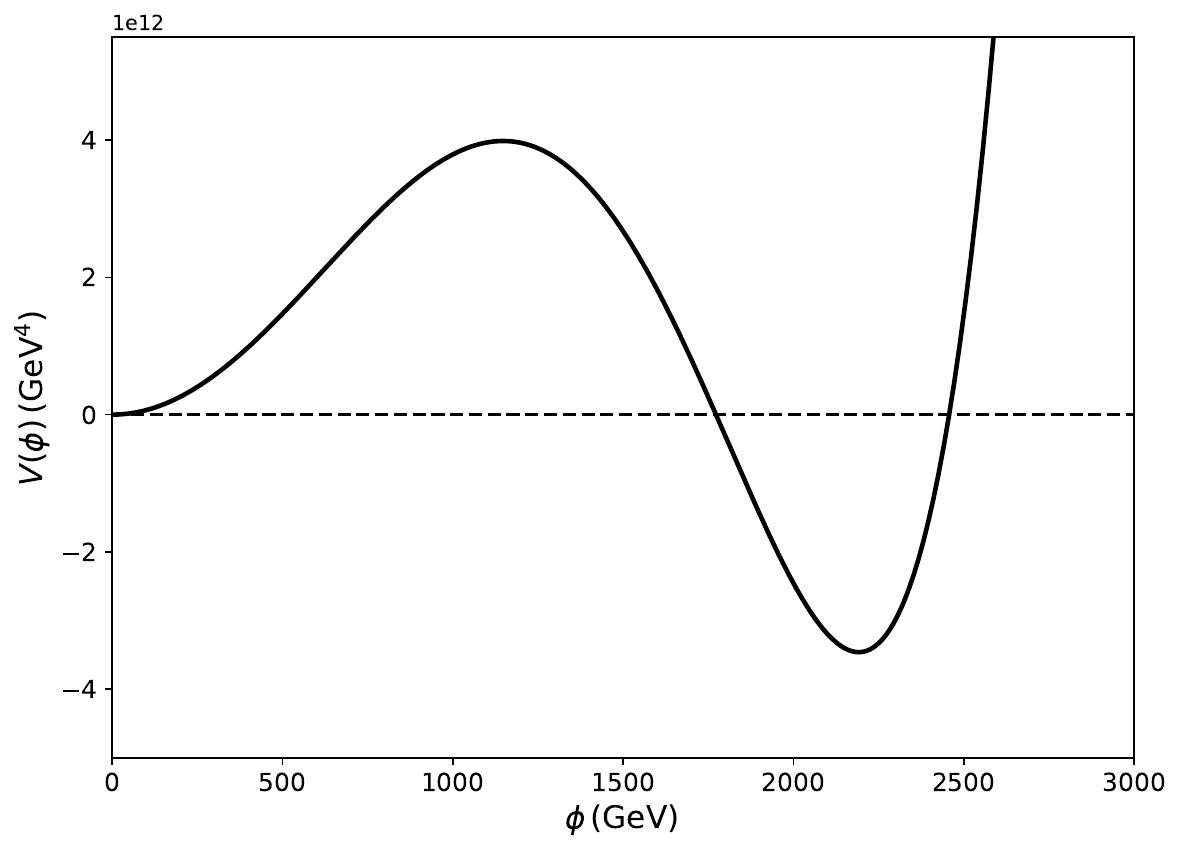}
\caption{This figure shows the potential $V(\phi)$ in Eq.~(\ref{Vpotential}). The true vacuum is at $\phi_{\text{true}} \approx 2191$ GeV. The parameters are chosen to be $\lambda=-3.227$, $\delta=1.867$, $m^2=6.667\times10^6$\,$\text{GeV}^2$, while the mass scale $\Lambda=2304$ GeV, which evades the collider constraints.}
\label{potential}
\end{figure}

We can see from the above figure that the potential possesses a stable true vacuum and supports a first-order phase transition by nucleating bubbles of the new vacuum within the old one.
By minimizing \(V(\phi)\) for \(\phi>0\) we locate the false vacuum at \(\phi=0\) (with \(V=0\)) and the true vacuum at
\begin{equation}
\phi_{\rm true}\approx 2191\;\mathrm{GeV},
\qquad
V(\phi_{\rm true})=-3.460\times10^{12}\;\mathrm{GeV}^4.
\end{equation}
Scanning between these two minima reveals a local maximum (the barrier) at
\begin{equation}
\phi_{\rm barrier}=1147\;\mathrm{GeV},
\qquad
V(\phi_{\rm barrier})=+3.987\times10^{12}\;\mathrm{GeV}^4.
\end{equation}
The energy difference is
\begin{equation}
\Delta V\equiv V_{\rm false}-V_{\rm true}
=3.460\times10^{12}\;\mathrm{GeV}^4,
\end{equation}
and the wall turning point \(\psi_0\) (first zero of \(V\)) occurs between the barrier and the true minimum. Integrating
\begin{equation}
S_1=\int_{0}^{\psi_0}d\phi\,\sqrt{2\,V(\phi)}
\end{equation}
yields
\begin{equation}
S_1=3.389\times10^9\;\mathrm{GeV}^3,
\end{equation}
and the thin-wall critical radius
\begin{equation}
R_0=\frac{3\,S_1}{\Delta V}
=2.939\times10^{-3}\;\mathrm{GeV}^{-1},
\end{equation}
which leads to an $O(4)$ bounce action
\begin{equation}
S_E \;=\; -\tfrac12\,\Delta V\,\pi^2\,R_0^4
\;+\;2\pi^2\,R_0^3\,S_1
\;=\;4.245\times10^2.
\end{equation}

The decay rate per unit volume per unit time is defined as $\Gamma \sim B e^{-S_E}$, where $B$ is a prefactor of order $v^4$. The requirement that our observable universe, with a four-volume of order $t_{\rm Hubble}^4$, remains in the unbroken phase corresponds to the condition
\begin{equation}
\Gamma t_{\rm Hubble}^4 \lesssim 1.
\end{equation}
Taking $t_{\rm Hubble}\sim 10^{10}$ years $\simeq 4.79\times10^{41}\,\mathrm{GeV}^{-1}$, we find
\begin{equation}
\log_{10}(\Gamma t_{\rm Hubble}^4) \;=\; -4.27\;\ll 0\;\;\Longrightarrow\;\;\Gamma t_{\rm Hubble}^4 \approx 10^{-4.27}\;\ll 1\,.
\end{equation}
This shows that, for the chosen values of the parameters, the false vacuum is long-lived on cosmological timescales, and most of the universe is still in the old (false) vacuum. At the same time, the suppression is not enormous, which leaves an interesting possibility that there might be a few bubbles here and there in the visible universe. 

In the false vacuum, the scalar field mass is
\begin{equation}
M_{\rm false}^2=\left.\frac{d^2V}{d\chi^2}\right|_{\phi=0}=m^2
\;\;\Longrightarrow\;\;
M_{\rm false}=2.58\times10^3\;\mathrm{GeV},
\end{equation}
while in the true vacuum its mass is
\begin{equation}
\mu^2
=\left.\frac{d^2V}{d\chi^2}\right|_{\chi=0}
=m^2+6\lambda\,v^2+\frac{15\delta}{\Lambda^2}\,v^4
\;=\;4\lambda\,v^2+\frac{12\delta}{\Lambda^2}\,v^4\;,
\qquad
\mu=5.942\times10^3\;\mathrm{GeV},
\end{equation}
for the benchmark parameters. Finally, with gauge coupling \(q=0.3\) we have a photon mass
\begin{equation}
m_\gamma
= q\,v
\;\;\Longrightarrow\;\;
m_\gamma\simeq 657\;\mathrm{GeV}.
\end{equation}

The findings in this section indicate that a phenomenologically viable metastable vacuum with a sufficiently high barrier, a controlled true-vacuum depth, and long enough lifetime can exist. Since the present Universe is assumed to remain in the false vacuum, collider constraints must be applied to the false-vacuum spectrum. For our benchmark, the scalar mass in the false vacuum is \(M_{\rm false}\simeq 2.58~\mathrm{TeV}\), which lies above the mass scales targeted by current direct scalar searches. A fully model-specific collider exclusion, however, would still depend on the scalar's production channels and decay pattern.

\section{Propagation of the true scalar vacuum bubble}

When the scalar field undergoes tunneling through its potential barrier, a bubble of true vacuum is generated and begins to expand. Neglecting interactions with other fields, and assuming that spacetime is approximately flat in the region of interest, the action for the scalar field reduces to
\begin{equation}
S(\Phi) = \int d^4x \left(\frac{1}{2}(\partial_\mu \Phi)^2 - V(\Phi)\right),
\end{equation}
which in turn leads to the classical equation of motion
\begin{equation}
-\partial_t^2 \Phi + \nabla^2 \Phi - V'(\Phi) = 0.
\end{equation}
After performing a Wick rotation, \(t \rightarrow i\tilde{\tau}\), the equation becomes
\begin{equation}
\partial_{\tilde{\tau}}^2 \Phi + \nabla^2 \Phi - V'(\Phi) = 0.
\end{equation}
For an \(O(4)\) symmetric bounce solution, the field depends only on the radial coordinate \(\rho = \sqrt{\tilde{\tau}^2 + r^2}\),  and the equation further simplifies to
\begin{equation}
\frac{d^2\Phi}{d\rho^2} + \frac{3}{\rho}\frac{d\Phi}{d\rho} = V'(\Phi).
\end{equation}
 The bounce solution that meets the boundary conditions at nucleation (i.e., at \(t=0\)) is given by
\begin{equation}
\Phi(t,\vec{x}) = \Phi\bigl(\rho=\sqrt{r^2-t^2}\bigr).
\end{equation}
In this formulation, the bubble contracts for \(t<0\), bounces at \(t=0\), and subsequently expands. Under the thin wall approximation, the scalar field configuration is modelled as
\begin{equation}
\Phi(\rho) = \begin{cases}
0, & \rho > R,\\[1ex]
v_1, & \rho < R,
\end{cases}
\end{equation}
where \(v\) and \(v_1\) denote the expectation values of the scalar field in the false and true vacuum states, respectively.

The creation of particles that accompany first-order phase transitions has been studied previously (e.g. \cite{Yamamoto:1994te,MersiniHoughton:1999tt,Mersini-Houghton:1999aoa,Maziashvili:2003sk,Maziashvili:2003zy,Vachaspati:1991tq,Swanson:1986hx,Hamazaki:1995dy,Maziashvili:2003kj,PhysRevD.19.1906,Espinosa:2010hh,Turner:1992tz,Quiros:1999jp,Affleck:1980ac,Steinhardt:1981ct}). To explicitly address particle production in Minkowski space, we assume homogeneous vacuum decay, as detailed in \cite{Maziashvili:2003kp,Tanaka:1993ez}.

\section{Particle production due to vacuum mismatch}

A difference between the false and true vacua generally results in particle production. In the present scenario, the false vacuum persists outside the bubble, while the inside assumes the true vacuum after tunneling. This vacuum discrepancy drives the creation of massive scalar particles. By decomposing the scalar field into a classical background and its fluctuation,
\[
\phi = v + \chi,
\]
the fluctuation field \(\chi\) satisfies
\begin{equation}
\partial_{\tilde{\tau}}^{2} \chi + \nabla^2 \chi - V''(\phi_c)\chi = 0.
\end{equation}
Neglecting any additional interactions with other fields, the above equation is approximated as
\begin{eqnarray}
\partial_{\tilde{\tau}}^2 \chi + \nabla^2 \chi - M^2\chi &=& 0,\quad \text{, for ${\tilde{\tau}}<\tilde{\tau}^*$}\\\\[1ex]
\partial_{\tilde{\tau}}^2 \chi + \nabla^2 \chi - \mu^2\chi &=& 0,\quad \text{, for ${\tilde{\tau}}>\tilde{\tau}^*$}\\,
\end{eqnarray}
where $\tilde{\tau}$ is the characteristic (Euclidean) time scale of the phase transition. 
We set $\tilde{\tau}=-R_0$, where $R_0$ denotes the bubble’s radius at nucleation. This choice reflects the fact that the bubble’s expansion is characterized by constant proper acceleration (as opposed to the coordinate acceleration, which varies). The trajectory of the bubble wall follows the hyperbolic equation of motion $r^2-t^2=R_0^2$, from which we derive the proper acceleration magnitude as 
$a=1/R_0$ (see Appendix~\ref{App-C}). This constant proper acceleration implies that particle production during bubble expansion is fundamentally tied to the Unruh effect, where the radiation spectrum is determined by the acceleration scale. Because the magnitude of the proper acceleration $a=1/R_0$ persists throughout the expansion, particle creation occurs continuously rather than as a transient effect.
It is crucial to distinguish between proper acceleration—defined in the momentarily comoving inertial frame of the bubble wall—and the coordinate acceleration observed from a fixed reference frame. The former governs local physical effects, such as Unruh radiation, while the latter depends on the observer’s frame and does not directly determine particle production. In fact, the coordinate acceleration approaches zero as the bubble wall approaches the speed of light.

 The solution for \(\chi\) can be expressed as a linear combination of mode functions \(g_k\) (which satisfy \(\nabla^2 g_k = -k^2\)):
\begin{eqnarray}
g_k=\left\{
  \begin{array}{lr}
   e^{\omega_- \tilde{\tau}} e^{i\vec{k}\cdot \vec{x}} &  \text{, for ${\tilde{\tau}}<\tilde{\tau}^*$}\\[1ex]
   A_k\, e^{\omega_+ \tilde{\tau}} e^{i\vec{k}\cdot \vec{x}} + B_k\, e^{-\omega_+ \tilde{\tau}} e^{i\vec{k}\cdot \vec{x}} & \text{, for ${\tilde{\tau}}>\tilde{\tau}^*$}.
  \end{array}
\right.
\end{eqnarray}
Here, \(\omega_+ = \sqrt{\mu^2+k^2}\) and \(\omega_- = \sqrt{M_{\rm false}^2+k^2}\), with \(M_{\rm false} = \sqrt{m^2}\) and \(\mu = \sqrt{m_\chi^2}\) representing the masses of the scalar field in the false and true vacua, respectively. In particular, the scalar mass in the false vacuum region is
\begin{equation}
M_{\rm false}^2 = m^2 \;>\; 0 \quad \text{for the benchmark } m^2>0.
\end{equation}
In the true vacuum region the scalar mass is
\begin{equation}
\mu^{2}
\;=\;
m_\chi^{2}
\;=\;
4\lambda v^2+\frac{12\delta v^4}{\Lambda^2}.
\end{equation}
Since both \(g_k\) and its derivative \(\partial_{\tilde{\tau}} g_k\) must remain continuous at 
\(\tilde{\tau}=\tilde{\tau}^*\), the coefficients \(A_k\) and \(B_k\) are determined to be
\begin{eqnarray}
A_k &=& \frac{1}{2\omega_+}(\omega_++\omega_-)e^{-(\omega_+-\omega_-)\tilde{\tau}^*},\\[1ex]
B_k &=& \frac{1}{2\omega_+}(\omega_+-\omega_-)e^{(\omega_++\omega_-)\tilde{\tau}^*}.
\end{eqnarray}

The particle creation spectrum is obtained from the Bogoliubov transform \cite{Tanaka:1993ez}
\begin{eqnarray}
N_k=\frac{B_k^2}{A_k^2-B_k^2} =  \left[ \frac{(\omega_++\omega_-)^2}{(\omega_+-\omega_-)^2}e^{4\omega_+ R_0}- 1 \right]^{-1}.
\label{pn}
\end{eqnarray}
\begin{figure}[t]
\begin{center}
\includegraphics[scale=0.28]{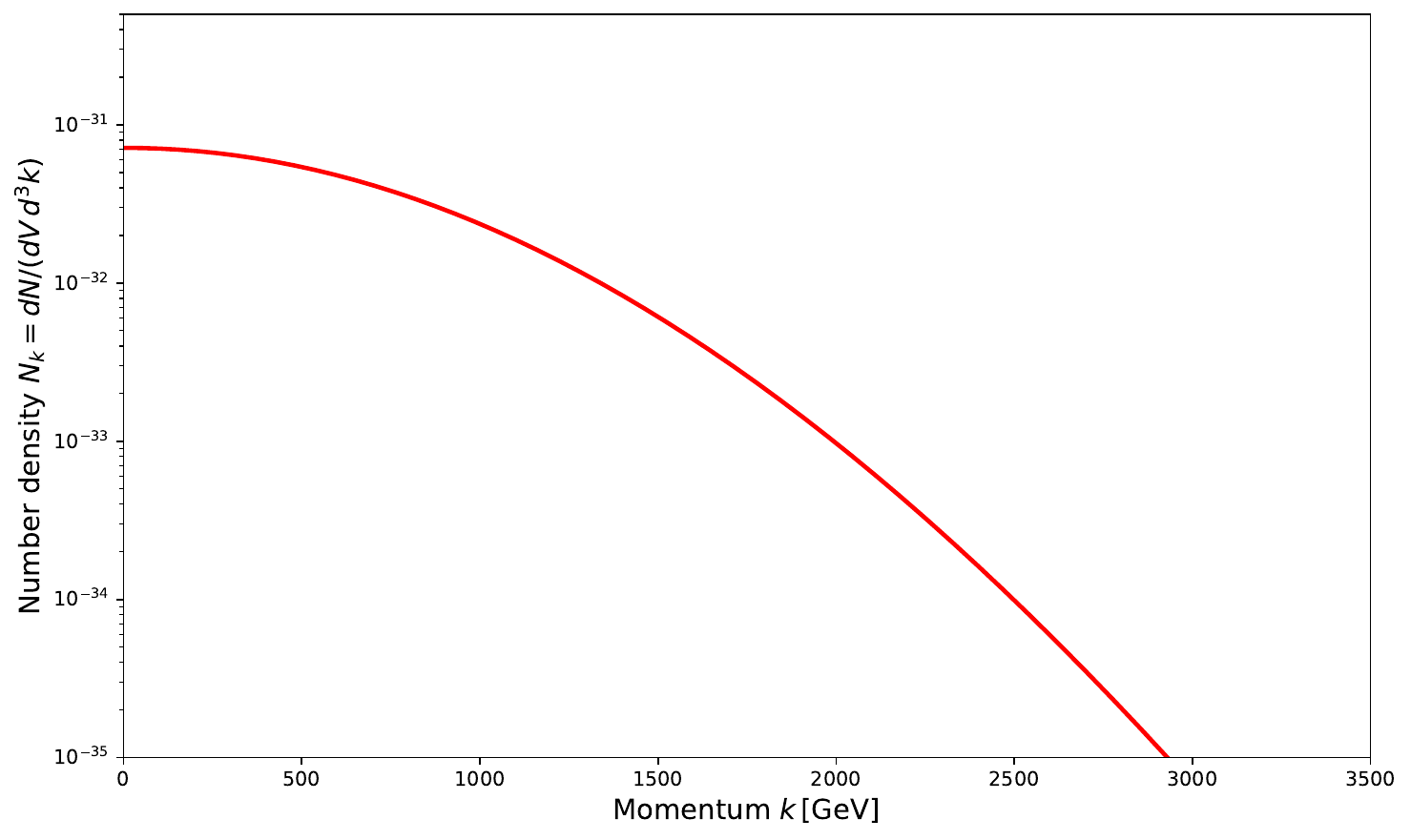}
\includegraphics[scale=0.28]{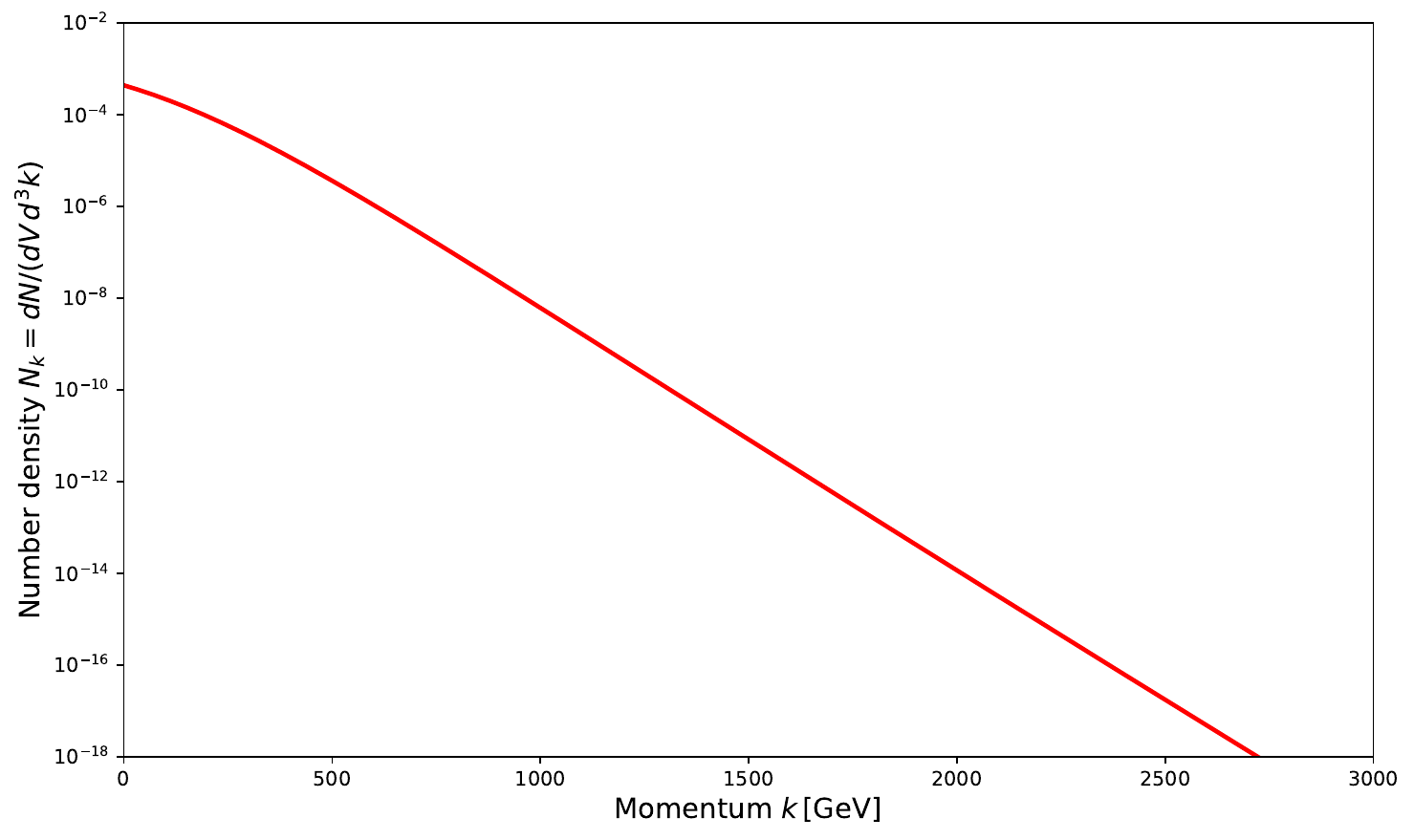}
\end{center}
\caption{\label{CSPlot}On the left: momentum-dependent occupation number $N_k$ of the scalar. On the right: momentum-dependent occupation number $n_k$ of the massive photon created due to the vacuum mismatch.}
\end{figure}
The particle creation spectra are derived using the Bogoliubov approach \cite{Tanaka:1993ez}. The left side of Figure~\ref{CSPlot} presents the momentum-dependent occupation number $N_k$ for the scalar field $\Phi$. More precisely, $N_k$ is the number density per unit phase volume, i.e. $N_k = dN/dVd\vec{k}$. The mass of the scalar in the false vacuum is set as \(M_{\rm false} = 2.58\times10^3\) GeV, the bubble radius is approximately \(R_0 \approx 2.94\times 10^{-3}\, \text{GeV}^{-1}\), and the scalar mass in the true vacuum is \(\mu = 5.94\times10^3\)\,GeV.

The right side of Figure~\ref{CSPlot} presents the momentum-dependent occupation number $n_k$ for the massive photon. In a scenario where the electromagnetic U(1)\(_{\rm EM}\) gauge symmetry is spontaneously broken at late times by a scalar acquiring a vacuum expectation value of order $v\simeq 2191$\,GeV and gauge charge \(q=0.3\), the photon field \(A_\mu\) absorbs the would-be Goldstone boson and becomes a massive Proca field inside the symmetry-broken region. Inside the bubble (${\tilde{\tau}}>\tilde{\tau}^*$) it acquires a mass
\begin{equation}
m_A \equiv m_\gamma = q\,v \simeq 657\;\mathrm{GeV},
\qquad
\tilde\tau = -R_0.
\end{equation}
Decomposing each physical polarization into momentum modes satisfying
\begin{equation}
A_k''(\tilde{\tau}) + \bigl[k^2 + m^2(\tilde{\tau})\bigr]\,A_k(\tilde{\tau}) = 0,
\quad
m(\tau)=
\begin{cases}
0, & \text{, for ${\tilde{\tau}}<\tilde{\tau}^*$}\\
m_A, & \text{, for ${\tilde{\tau}}>\tilde{\tau}^*$},
\end{cases}
\end{equation}
and matching at \(\tilde{\tau}=\tilde{\tau}^*\) yields the Bogoliubov coefficients
\begin{equation}
\alpha_k = \frac{\omega_+ + \omega_-}{2\omega_+}e^{-i(\omega_+ - \omega_-)\tilde{\tau}^*},
\qquad
\beta_k = \frac{\omega_+ - \omega_-}{2\omega_+}e^{-i(\omega_+ + \omega_-)\tilde{\tau}^*},
\end{equation}
with
\(\omega_-=\sqrt{k^2}\), \(\omega_+=\sqrt{k^2+m_A^2}\).  The occupation number per mode is
\begin{equation}
N_k
= \frac{|\beta_k|^2}{|\alpha_k|^2-|\beta_k|^2}
= \Bigl[\tfrac{(\omega_+ + \omega_-)^2}{(\omega_+ - \omega_-)^2}
\,e^{4\,\omega_+\,R_0}-1\Bigr]^{-1}.
\end{equation}
At zero comoving momentum the matching behaves differently for transverse and longitudinal polarizations. For the two \emph{transverse} modes the unbroken phase has a well-defined massless photon with frequency
\(\omega_-=|k|=0\), while in the broken phase the Proca field has
\(\omega_+=\sqrt{k^2+m_A^2}=m_A\).
Using the standard mass-quench matching, the Bogoliubov occupation number reduces to
\begin{equation}
N^{(T)}_{k=0} \;=\; \Bigl[e^{\,4 m_A R_0}-1\Bigr]^{-1},
\end{equation}
which is perfectly well defined for each transverse polarization. A conservative, gauge-independent count at $k=0$ therefore gives
\begin{equation}
n_{\text{photon}}(k=0)\;=\;2\,N^{(T)}_{k=0}
\;=\;\frac{2}{e^{\,4 m_A R_0}-1}\,.
\end{equation}

The longitudinal polarization is subtler. In the unbroken phase there is no longitudinal photon; this degree of freedom is created only after symmetry breaking when the would-be Goldstone is eaten. Consequently, the simple two-oscillator matching does not apply. A proper treatment requires working with the coupled \((A_\mu,\theta)\) system and matching the gauge-invariant combination
\begin{equation}
\partial_\mu \theta \;-\; q\,v\,A_\mu\,,
\end{equation}
which mixes the scalar phase with the vector field. The resulting longitudinal Bogoliubov coefficient generally differs from the transverse one. A complete derivation of the longitudinal spectrum from the coupled scalar--vector dynamics is beyond the scope of this work and will be presented elsewhere. For the purposes of the present phenomenology, we therefore include only the two transverse contributions at $k=0$; the longitudinal mode can be added once the coupled analysis is performed.

\section{Relativistic bubble wall dynamics in a viscous medium and terminal velocity}
\label{sec:bubble-dynamics-u1em}

If a bubble wall were to move at the speed of light, no signal could outrun it to forewarn us of its arrival. In practice, however, the growth of a vacuum bubble is impeded by its interactions with the ambient medium---whether this consists of matter, radiation, or excitations generated by the bubble itself. Consequently, the wall does not accelerate indefinitely but instead approaches a finite terminal velocity below the speed of light. In this work we apply the relativistic thin--wall formalism developed previously for Higgs vacuum decay with dissipative effects~\cite{Sengupta:2025jah}, and extend it to the present $U(1)_{EM}$ symmetry--breaking case (see also~\cite{Moore_2000, Moore:1995ua, PhysRevD.108.103523, Gouttenoire:2021kjv, Bodeker:2017cim, Megevand:2009gh} for related discussions of wall friction and hydrodynamics). For completeness we collect the relevant dynamical equations and specify the $U(1)_{EM}$ model--specific parameters.

\subsection{Equation of motion and terminal balance}
\label{subsec:eom-u1em}

In the thin--wall picture the rest energy per unit area of the bubble is the surface tension $\sigma$. Under a Lorentz boost with velocity $v$, the wall energy density becomes $\sigma\gamma(v)$ and the momentum density $\sigma\gamma(v)v$, where $\gamma(v)=(1-v^2)^{-1/2}$. Balancing the time derivative of this momentum density against the driving pressure, curvature pressure, and dissipative drag yields the relativistic spherical equation of motion~\cite{Sengupta:2025jah}:
\begin{equation}
  \sigma\,\gamma^3(v)\,\frac{dv}{dt}
  \;=\;
  \Delta V \;-\; \frac{2\sigma}{R(t)} \;-\; \eta\,\gamma(v)\,v,
  \label{eq:spherical_EOM_u1em}
\end{equation}
where $R(t)$ is the instantaneous bubble radius, $\Delta V \equiv V_{\rm false}-V_{\rm true}$ is the latent--heat pressure, and $\eta$ encodes the effective linear friction coefficient describing wall–medium scattering.

At terminal motion, where acceleration vanishes, Eq.~\eqref{eq:spherical_EOM_u1em} simplifies to
\begin{equation}
  0 \;=\; \Delta V \;-\; \frac{2\sigma}{R} \;-\; \eta\,\gamma(v)\,v
  \qquad\Longleftrightarrow\qquad
  \Delta V_{\rm eff}(R)\;\equiv\;\Delta V-\frac{2\sigma}{R}
  \;=\; \eta\,\gamma(v)\,v,
  \label{eq:terminal_balance_u1em}
\end{equation}
so that in the planar limit $R\to\infty$ one obtains $\Delta V=\eta\,\gamma v$.

\medskip
For the $U(1)_{EM}$ scalar potential given in Sec.~\ref{Model} and Eq.\eqref{Vpotential}, the inputs to Eq.~\eqref{eq:spherical_EOM_u1em} are
\begin{equation}
  \Delta V \;=\; V_{\rm false}-V_{\rm true}, 
  \qquad
  \sigma \;=\; S_1,
\end{equation}
where $S_1=\int_0^{\psi_0} d\phi\,\sqrt{2V(\phi)}$ is the surface tension. One finds
\begin{equation}
  A \;\equiv\; \frac{\Delta V}{\sigma}
  \;=\; \frac{3}{R_0},
  \label{eq:A_equals_3_over_R0_u1em}
\end{equation}
with $R_0$ the thin--wall nucleation radius. This sets the initial drive.

\subsection{Proper--time formulation}
\label{subsec:proper-u1em}

It is useful to parametrize the dynamics by the wall’s proper time $\tau$ and rapidity $y(\tau)$:
\begin{equation}
  v=\tanh y,\qquad
  \gamma=\cosh y,\qquad
  \gamma v=\sinh y,\qquad
  \frac{dt}{d\tau}=\gamma,\qquad
  \frac{dR}{d\tau}=\sinh y.
  \label{eq:kinematics_u1em}
\end{equation}
The invariant acceleration is
\begin{equation}
  \alpha(\tau)\;\equiv\;\gamma^3\,\frac{dv}{dt}\;=\;\frac{dy}{d\tau}.
  \label{eq:alpha_def_u1em}
\end{equation}
Dividing Eq.~\eqref{eq:spherical_EOM_u1em} by $\sigma$ and using \eqref{eq:alpha_def_u1em} gives
\begin{align}
  \frac{dy}{d\tau}
  &= A \;-\; \frac{2}{R(\tau)} \;-\; B\,\sinh y(\tau),
  \qquad
  A=\frac{\Delta V}{\sigma},\ \ B=\frac{\eta}{\sigma},
  \label{eq:dy_dtau_u1em}\\
  \frac{dt}{d\tau}&=\cosh y,\qquad
  \frac{dR}{d\tau}=\sinh y.
  \label{eq:dt_dR_dtau_u1em}
\end{align}
With $R(0)=R_0$ and $y(0)=0$, the initial proper acceleration is
\begin{equation}
  \alpha(0)=A-\frac{2}{R_0}\;\simeq\;\frac{1}{R_0},
  \label{eq:alpha0_u1em}
\end{equation}
where $A=3/R_0$ has been used. For $\tau\ll R_0$, $y(\tau)=\alpha(0)\tau+\mathcal{O}(\tau^2)$ and $\sinh y\simeq y$, so that
\begin{equation}
  \sinh y(\tau)\simeq \alpha(0)\tau,
  \qquad
  R(\tau)\simeq R_0+\tfrac{\alpha(0)}{2}\tau^2.
  \label{eq:early_kin_u1em}
\end{equation}
Inserting into \eqref{eq:dy_dtau_u1em} yields the small--$\tau$ series
\begin{equation}
  \alpha(\tau)\simeq \frac{1}{R_0}-\frac{B}{R_0}\tau+\frac{\tau^2}{R_0^3}
  +\mathcal{O}(\tau^3),
  \label{eq:alpha_series_u1em}
\end{equation}
valid for $B\,R_0\ll1$. The linear term arises from viscous drag, the quadratic from curvature relaxation.

\subsection{Evolution of the proper acceleration and terminal motion}
\label{subsec:alpha-evolution-u1em}

Differentiating Eq.~\eqref{eq:dy_dtau_u1em} and using $\dot R=\sinh y$ gives
\begin{equation}
  \frac{d\alpha}{d\tau}
  = \frac{2\,\sinh y}{R^2}-B\cosh y\,\alpha.
  \label{eq:alpha_prime_u1em}
\end{equation}
In the planar, small--rapidity limit this reduces to $d\alpha/d\tau\simeq -B\alpha$, giving
\begin{equation}
  \alpha(\tau)\simeq \alpha(0)e^{-B\tau},\qquad
  \tau_{\rm term}=\frac{\sigma}{\eta}.
  \label{eq:tau_term_u1em}
\end{equation}
At terminal balance ($\alpha=0$), the rapidity satisfies
\begin{equation}
  \sinh y_{\rm term}=\frac{A}{B}
  \qquad\Longleftrightarrow\qquad
  \gamma_{\rm term}v_{\rm term}=\frac{\Delta V}{\eta}.
  \label{eq:terminal_sinh_u1em}
\end{equation}

\section{Particle production due to vacuum mismatch in presence of friction}
\label{sec:u1em-prod-spherical}

In this section we estimate the number of quanta produced before the bubble wall saturates at its terminal velocity. For clarity, we work in GeV units. The relevant benchmark parameters are
\begin{equation}\nonumber
v=2191\;\text{GeV},\qquad
\Delta V=3.460\times10^{12}\;\text{GeV}^4,\qquad
\sigma=S_1=3.389\times10^{9}\;\text{GeV}^3,
\end{equation}
\begin{equation}
R_0=2.939\times10^{-3}\;\text{GeV}^{-1},\qquad
A=\frac{\Delta V}{\sigma}=1.021\times10^{3}\;\text{GeV},\qquad
\mu=5.942\times10^{3}\;\text{GeV},
\label{eq:inputs-u1em}
\end{equation}
together with the false-vacuum scalar mass
\begin{equation}
M_{\rm false}=2.582\times10^{3}\;\text{GeV},
\end{equation}
and the broken-phase photon mass
\begin{equation}
m_\gamma = 657\;\text{GeV}.
\end{equation}

The dimensionless ratios are
\begin{equation}
A=\frac{\Delta V}{\sigma},\qquad B=\frac{\eta}{\sigma},
\end{equation}
and the wall kinematics are parametrized as
\begin{equation}
v=\tanh y,\quad \gamma=\cosh y,\quad \gamma v=\sinh y,\quad
\frac{dt}{d\tau}=\gamma,\quad \frac{dR}{d\tau}=\sinh y.
\label{Parameters-u1em}
\end{equation}
The proper acceleration is
\begin{equation}
\alpha(\tau)=\frac{dy}{d\tau}=A-\frac{2}{R(\tau)}-B\sinh y(\tau),
\label{eq:alpha-spherical-u1em}
\end{equation}
with initial conditions
\begin{equation}
R(0)=R_0,\qquad y(0)=0,\qquad t(0)=0,\qquad N_{\rm tot}(0)=0.
\end{equation}
At nucleation one has \(\alpha(0)\simeq 1/R_0\), using \(A\simeq 3/R_0\).

Replacing the constant proper acceleration \(a=1/R_0\) in the vacuum-mismatch expression by the time-dependent acceleration in Eq.~\eqref{eq:alpha-spherical-u1em}, the instantaneous zero-momentum occupation number becomes
\begin{equation}
N_{k=0}(\tau)=\left[\frac{(\omega_++\omega_-)^2}{(\omega_+-\omega_-)^2}
\exp\!\left(\frac{4\omega_+}{\alpha(\tau)}\right)-1\right]^{-1}.
\label{eq:Nk0-alpha-u1em}
\end{equation}
For the scalar we use
\begin{equation}
\omega_+=\mu,\qquad \omega_-=M_{\rm false},
\end{equation}
while for the massive photon
\begin{equation}
\omega_+=m_\gamma,\qquad \omega_-=0.
\end{equation}
The accumulated number then evolves as
\begin{equation}
\frac{dN_{\rm tot}}{d\tau}=N_{k=0}(\tau)\,4\pi R(\tau)^2\sinh y(\tau).
\label{eq:dNtot-u1em}
\end{equation}

The details of the numerical calculations are shown in the Appendix~\ref{App-D}. The resulting integrated yields are summarized in Table~\ref{FrictionalYield}. For each benchmark deficit \(\delta\), we evolve the wall until \(\tau_{\rm term}\), and quote the accumulated particle number from the \(k=0\) mode only. The scalar yield is shown for a single scalar degree of freedom, while the massive-photon yield corresponds to the two transverse photon modes as a conservative estimate. Inclusion of the longitudinal photon mode would increase the total multiplicities, but does not change the qualitative hierarchy between the channels.

\begin{table*}[tbhp]
\centering
\small
\begin{tabular}{@{}lccccc@{}}
\toprule
Scenario & $\eta\,[{\rm GeV}^4]$ & $\tau_{\rm term}\,[{\rm GeV}^{-1}]$ & $R_{\rm fin}\,[{\rm GeV}^{-1}]$ & Scalar--$N_{\rm tot}^{\rm(int)}$ & Massive Photon--$N_{\rm tot}^{\rm(int)}$ \\
\midrule
$\delta=10^{-12}$ & $4.893\times10^{6}$ & $6.927\times10^{2}$ & $4.898\times10^{8}$ & $6.295\times10^{-6}$ & $2.001\times10^{7}$ \\
$\delta=10^{-11}$ & $1.547\times10^{7}$ & $2.190\times10^{2}$ & $4.898\times10^{7}$ & $1.989\times10^{-7}$ & $6.328\times10^{5}$ \\
$\delta=10^{-10}$ & $4.893\times10^{7}$ & $6.927\times10^{1}$ & $4.897\times10^{6}$ & $6.274\times10^{-9}$ & $2.003\times10^{4}$ \\
\bottomrule
\end{tabular}
\caption{For each terminal deficit from the speed of light, \(\delta=1-v_{\rm term}\), we evolve the \(U(1)_{EM}\) bubble wall using Eq.~\eqref{eq:alpha-spherical-u1em} up to \(\tau_{\rm term}=\sigma/\eta\). The integrated yields \(N_{\rm tot}^{\rm(int)}\) are shown separately for the scalar and for the massive photon (only transverse modes included). Only the \(k=0\) mode is included; higher-momentum modes would further enhance the total production.}
\label{FrictionalYield}
\end{table*}

The table shows that as the wall expands, friction gradually compensates the vacuum-pressure drive and the motion approaches terminal velocity on the timescale \(\tau_{\rm term}\). The final yield is governed by a competition between the exponential suppression of \(N_{k=0}\) as the acceleration decreases and the rapid growth of the geometric factor \(4\pi R^2\sinh y\). In the present \(U(1)_{EM}\) benchmark this competition strongly favors the massive-photon channel, while the scalar channel is heavily suppressed because of the much larger scalar mass in the true vacuum.

The produced excitations are massive photons and scalar quanta generated near the accelerating bubble wall. These unstable modes subsequently decay into Standard Model particles and may source energetic photons and neutrinos. In particular, the photon sector directly reflects the fact that the gauge boson acquires a mass during \(U(1)_{EM}\) breaking, while the scalar sector probes the curvature of the effective potential around the true minimum. Since the scalar mass threshold is much higher than the photon mass threshold, scalar production is exponentially suppressed over the entire benchmark range considered here.

For completeness, we note that including modes with \(k>0\) would increase the total multiplicity by an overall factor set by the acceleration scale and the relevant mass thresholds, but this would not modify the qualitative features of the spectra or the conclusions. In particular, for ultra-relativistic walls (\(\delta\ll1\)) the integrated massive-photon yield is significantly enhanced by the large bubble radius reached before \(\tau_{\rm term}\), whereas stronger friction suppresses the production by driving the wall more rapidly into the terminal regime. The resulting differential number densities, \(dN_\gamma/(dE\,dV)\) and \(dN_\nu/(dE\,dV)\), therefore remain the key observables for comparing different nucleation scenarios in the broken-\(U(1)_{EM}\) phase.

Particles can also be produced through scattering processes off the bubble wall once it enters the steady-state regime, providing another possible source of heavy broken-phase excitations and their subsequent photon and neutrino decay products; however, this mechanism is not expected to dominate, since it is suppressed relative to the much larger thermal production generated by frictional dissipation behind the wall. Therefore, we discuss the particle production due to thermal dissipation in the following section.

\section{Thermal particle production from frictional dissipation in the \texorpdfstring{$U(1)_{\rm EM}$}{U(1)EM} transition}
\label{sec:thermal-u1em}

As the $U(1)_{\rm EM}$-breaking bubble wall propagates through an ambient medium, microscopic scatterings with the surrounding plasma exert a frictional pressure
\begin{equation}
P_{\rm fric}=\eta\,\gamma v,
\end{equation}
which opposes the vacuum pressure driving the wall outward. In the absence of friction the wall would continue to gain kinetic energy, whereas in the physical case a fraction of this energy is dissipated into a thin shocked layer behind the wall. Because the wall is ultra-relativistic, this layer thermalises on timescales much shorter than the macroscopic evolution time of the bubble. Our goal in this section is to estimate the associated thermal energy deposition and the resulting production of heavy quanta in the broken $U(1)_{\rm EM}$ phase.

Our treatment follows the same energy-deficit logic used in our previous study of false Higgs vacuum decay \cite{Sengupta:2025jah} and in the $SU(3)_c$ analysis \cite{Sengupta:2025cdm}, but is now adapted to the present $U(1)_{\rm EM}$ setup. We do not assume a detailed microscopic model for the friction; instead, dissipative effects are encoded phenomenologically through the difference between a frictionless bubble trajectory and the corresponding friction-limited one, which is then converted into local heating of the shocked shell.

\subsection*{Energy deficit and local heating}

The boosted wall energy per unit area is
\begin{equation}
E_{\rm wall}(t)=\sigma\,\gamma(t),
\qquad
\gamma(t)=\frac{1}{\sqrt{1-v^2(t)}}\,,
\end{equation}
where $\sigma$ is the wall surface tension and $v(t)$ is the wall velocity in the rest frame of the surrounding medium. Curvature contributes through the usual Laplace pressure term $2\sigma/R$, but does not alter the form of the boosted surface energy.

To track dissipation we compare two trajectories:
\begin{itemize}
\item[(i)] a frictionless trajectory with velocity $v_0(t)$ and Lorentz factor $\gamma_0(t)$,
\item[(ii)] the physical friction-limited trajectory with velocity $v(t)$ and Lorentz factor $\gamma(t)$.
\end{itemize}
The frictionless wall obeys
\begin{equation}
\sigma\,\gamma_0^3\,\frac{dv_0}{dt}
=
\Delta V-\frac{2\sigma}{R},
\label{eq:u1-frictionless-eom}
\end{equation}
while the physical wall satisfies
\begin{equation}
\sigma\,\gamma^3\,\frac{dv}{dt}
=
\Delta V-\frac{2\sigma}{R}-\eta\,\gamma v.
\label{eq:u1-friction-eom}
\end{equation}
Using
\begin{equation}
\frac{d\gamma}{dt}=\gamma^3 v\,\frac{dv}{dt},
\end{equation}
one finds
\begin{equation}
\frac{d\gamma_0}{dt}
=
\frac{v_0}{\sigma}\left(\Delta V-\frac{2\sigma}{R}\right),
\qquad
\frac{d\gamma}{dt}
=
\frac{v}{\sigma}\left(\Delta V-\frac{2\sigma}{R}-\eta\,\gamma v\right).
\end{equation}
The maximal wall energy per unit area attainable in the absence of friction is
\begin{equation}
E_{\rm max}(t)=\sigma\,\gamma_0(t),
\end{equation}
so the energy deficit is
\begin{equation}
\Delta E(t)=\sigma\,[\gamma_0(t)-\gamma(t)].
\end{equation}
Differentiating gives
\begin{equation}
\frac{d}{dt}\Delta E(t)
=
(v_0-v)\left(\Delta V-\frac{2\sigma}{R}\right)
+\eta\,\gamma v^2.
\label{eq:u1-dDeltaE-master}
\end{equation}
This quantity measures the energy continuously extracted from the wall and deposited into the surrounding shocked shell.

We model the heated region as a comoving layer of thickness $\ell$ and area $A(t)=4\pi R^2(t)$, so that
\begin{equation}
E_{\rm th}(t)=\rho_{\rm th}(t)\,A(t)\,\ell,
\qquad
\rho_{\rm th}(t)\,\ell=\Delta E(t).
\end{equation}
In terms of proper time $d\tau=dt/\gamma$,
\begin{equation}
\frac{d\rho_{\rm th}}{d\tau}
=
\frac{\gamma}{\ell}
\left[
(v_0-v)\left(\Delta V-\frac{2\sigma}{R}\right)
+\eta\,\gamma v^2
\right].
\label{eq:u1-rho-general}
\end{equation}

\subsection*{Scalar and massive-photon channels}

In the broken $U(1)_{\rm EM}$ phase the heavy modes relevant for our analysis are the neutral radial scalar and the massive photon. For the benchmark introduced earlier,
\begin{equation}
\mu_s \equiv m_\Phi = 5.942~{\rm TeV},
\qquad
\mu_\gamma \equiv m_\gamma = 0.657~{\rm TeV}.
\end{equation}
As in the $SU(3)_c$ case, we assume that the effective thermalisation thickness in each channel is set by the inverse mass scale of the corresponding heavy mode,
\begin{equation}
\ell_s\sim \mu_s^{-1},
\qquad
\ell_\gamma\sim \mu_\gamma^{-1}.
\end{equation}
We therefore introduce separate energy densities $\rho_s$ and $\rho_\gamma$ obeying
\begin{equation}
\frac{d\rho_i}{d\tau}
=
\mu_i
\left[
\gamma (v_0-v)\left(\Delta V-\frac{2\sigma}{R}\right)
+\eta\,\gamma^2 v^2
\right],
\qquad
i=s,\gamma.
\label{eq:u1-rho-twochannel}
\end{equation}

The corresponding temperatures are
\begin{equation}
T_i(\tau)
=
\left(
\frac{30}{\pi^2 g_*}\,\rho_i(\tau)
\right)^{1/4},
\qquad i=s,\gamma,
\end{equation}
with $g_*=106.75$. The equilibrium number densities are then taken to be
\begin{equation}
n_i(\tau)
=
\frac{\zeta(3)}{\pi^2}\,g_i\,T_i^3(\tau),
\qquad
g_s=1,\qquad g_\gamma=2,
\label{eq:u1-numberdensities}
\end{equation}
where $g_\gamma=2$ corresponds to retaining only the two transverse polarizations of the massive photon, while the longitudinal mode is omitted as a conservative lower bound. For the benchmark considered here, the wall remains ultra-relativistic throughout the relevant stage of the evolution, and the shock temperatures obtained numerically are comfortably above the masses of both heavy species, $T_s \gg \mu_s$ and $T_\gamma \gg \mu_\gamma.$ Thus the thermally produced scalars and massive photons are themselves highly relativistic in the shocked layer. In this regime, using the standard relativistic equilibrium scaling $n_i \propto T_i^3$ is well justified for both sectors.

As the wall expands, it sweeps out a comoving volume
\begin{equation}
dV=4\pi R^2(\tau)\,\sinh y(\tau)\,d\tau,
\end{equation}
so the thermal production rates are
\begin{align}
\frac{dN_s}{d\tau}
&=
4\pi R^2(\tau)\,\sinh y(\tau)\,n_s(\tau),\\[1ex]
\frac{dN_\gamma}{d\tau}
&=
4\pi R^2(\tau)\,\sinh y(\tau)\,n_\gamma(\tau).
\label{eq:u1-dNdt}
\end{align}
The total multiplicities are obtained at the end of the evolution,
\begin{equation}
N_s=N_s(\tau_{\rm final}),
\qquad
N_\gamma=N_\gamma(\tau_{\rm final}).
\end{equation}

\subsection*{Numerical setup}

For the $U(1)_{\rm EM}$ benchmark we use
\begin{equation}
\Delta V = 3.460~{\rm TeV}^4,
\qquad
\sigma = 3.389~{\rm TeV}^3,
\qquad
R_0 = 2.939~{\rm TeV}^{-1},
\end{equation}
together with
\begin{equation}
\mu_s = 5.942~{\rm TeV},
\qquad
\mu_\gamma = 0.657~{\rm TeV}.
\end{equation}
We study three ultra-relativistic terminal-velocity deficits,
\begin{equation}
\delta = 10^{-12},\qquad 10^{-11},\qquad 10^{-10},
\end{equation}
with
\begin{equation}
v_{\rm term}=1-\delta,
\qquad
\gamma_{\rm term}=\frac{1}{\sqrt{1-v_{\rm term}^2}}.
\end{equation}
In the constant-friction approximation the force-balance condition gives
\begin{equation}
\eta(0)=\frac{\Delta V}{\gamma_{\rm term}v_{\rm term}}.
\label{eq:u1-etarget}
\end{equation}
To mimic the rise of drag in the heated shell, we promote the friction coefficient to
\begin{equation}
\eta(\tau)=g_{\rm eff}^2\,T_{\rm eff}^4(\tau),
\qquad
T_{\rm eff}^4(\tau)=T_{\rm amb}^4+T_{\rm shock}^4(\tau),
\end{equation}
with
\begin{equation}
T_{\rm shock}^4(\tau)=\frac{30}{\pi^2 g_*}\,[\rho_s(\tau)+\rho_\gamma(\tau)],
\qquad
g_{\rm eff}=10^{-3}.
\end{equation}
This ansatz is also physically motivated from simple kinetic-theory considerations. The frictional drag is set by the momentum flux of the thermal particles impinging on the wall, which scales as $n\,p \sim T^3\times T \sim T^4,$ multiplied by the interaction probability governing momentum transfer to the wall, parametrically of order $g_{\rm eff}^2.$ It is therefore natural that the drag term in the wall equation scales as $\eta(\tau)\gamma(\tau)v(\tau)\sim g_{\rm eff}^2 T_{\rm eff}^4,$ corresponding to an energy-loss rate with the same temperature dependence. Parametrically, this is consistent with more detailed analyses of ultra-relativistic electroweak bubble walls, where the thermal friction is likewise found to scale as $P_{\rm th}\sim \gamma^2 T^4,$ up to coupling-dependent factors \cite{Hoche:2020ysm}. The ambient temperature is fixed so that the initial drag matches Eq.~\eqref{eq:u1-etarget},
\begin{equation}
T_{\rm amb}
=
\left[\frac{\eta(0)}{g_{\rm eff}^2}\right]^{1/4}.
\end{equation}
The reference acceleration timescale is
\begin{equation}
\tau_{\rm term}=\frac{\sigma}{\eta(0)},
\end{equation}
and, we integrate up to
\begin{equation}
\tau_{\rm final}=5\,\tau_{\rm term}.
\end{equation}
This captures the dominant fraction of the energy deficit while keeping the evolution within the regime where our approximations remain reliable. (see Appendix~\ref{App-B} for details).

\subsection*{Numerical results for the \texorpdfstring{$U(1)_{\rm EM}$}{U(1)EM} benchmark}

The resulting thermal multiplicities and timescales are summarized in Table~\ref{tab:u1em-thermal-yields}.

\begin{table}[t]
\centering
\small
\begin{tabular}{@{}lcccc@{}}
\toprule
Scenario ($\delta$) 
& $\eta(0)\,[{\rm TeV}^4]$
& $\tau_{\rm final}\,[{\rm TeV}^{-1}]$
& Scalars $N_s$
& Massive photons $N_\gamma$ \\
\midrule
$10^{-12}$ &
$4.89\times10^{-6}$ &
$3.46\times10^{6}$ &
$6.84\times10^{24}$ &
$2.63\times10^{24}$ \\
$10^{-11}$ &
$1.55\times10^{-5}$ &
$1.10\times10^{6}$ &
$8.38\times10^{23}$ &
$3.21\times10^{23}$ \\
$10^{-10}$ &
$4.89\times10^{-5}$ &
$3.46\times10^{5}$ &
$9.87\times10^{22}$ &
$3.79\times10^{22}$ \\
\bottomrule
\end{tabular}
\caption{
Thermal particle yields in the scalar channel ($N_s$) and in the massive-photon channel ($N_\gamma$) obtained from the energy-deficit formulation of the $U(1)_{\rm EM}$ transition with temperature-dependent drag,
$\eta(\tau)=g_{\rm eff}^2 T_{\rm eff}^4(\tau)$ and $g_{\rm eff}=10^{-3}$.  
The massive-photon multiplicity includes only the two transverse polarizations. For each $\delta$, the reference drag $\eta(0)$ is fixed by the terminal condition $\Delta V=\eta(0)\gamma_{\rm term}v_{\rm term}$, and the evolution is integrated up to $\tau_{\rm final}=5\,\tau_{\rm term}$.
}
\label{tab:u1em-thermal-yields}
\end{table}

The scalar multiplicity exceeds the massive-photon multiplicity by a factor of a few over the benchmark range, reflecting the larger scalar heating scale $\mu_s$ in Eq.~\eqref{eq:u1-rho-twochannel}, even though the photon channel benefits from two polarization states. In total, the number of heavy quanta produced thermally lies in the range
\begin{equation}
N_s+N_\gamma \sim 10^{23}\text{--}10^{25},
\end{equation}
depending on the terminal-velocity deficit. As expected, smaller $\delta$ corresponds to a more ultra-relativistic wall, a longer acceleration time, a larger swept volume, and therefore substantially enhanced thermal production.

For the same runs, the final thermal energy densities are
\begin{align}
\rho_s &=
\left\{
4.01\times10^{7},\,
7.37\times10^{7},\,
1.44\times10^{8}
\right\}
~{\rm TeV}^4,
\\[1ex]
\rho_\gamma &=
\left\{
4.44\times10^{6},\,
8.16\times10^{6},\,
1.59\times10^{7}
\right\}
~{\rm TeV}^4,
\end{align}
for $\delta=\{10^{-10},10^{-11},10^{-12}\}$ respectively. These correspond to final channel temperatures
\begin{align}
T_s &=
\left\{
32.69,\,
38.06,\,
44.98
\right\}
~{\rm TeV},
\\[1ex]
T_\gamma &=
\left\{
18.85,\,
21.95,\,
25.94
\right\}
~{\rm TeV},
\end{align}
which are well above both $m_\Phi$ and $m_\gamma$. The thermally produced scalar and massive-photon quanta are therefore highly relativistic in the parameter range of interest.

\subsection{Thermal spectra for the scalar and massive-photon sectors}

Once the total energy densities and particle numbers are known, the momentum distributions follow from equilibrium thermodynamics. For a bosonic species of mass $m_i$ in a bath of temperature $T_i$, the massive Bose--Einstein spectral shape is
\begin{equation}
f_i(k)
=
\frac{k^2}{\exp\!\left(\sqrt{k^2+m_i^2}/T_i\right)-1},
\qquad
i=s,\gamma,
\label{eq:u1-massiveBE}
\end{equation}
with
\begin{equation}
m_s=\mu_s=5.942~{\rm TeV},
\qquad
m_\gamma=\mu_\gamma=0.657~{\rm TeV}.
\end{equation}
The physically normalized spectra are then
\begin{equation}
\frac{dN_i}{dk}
=
N_i\,
\frac{f_i(k)}{\int_0^\infty f_i(k)\,dk},
\qquad i=s,\gamma,
\label{eq:u1-normalised-spectrum}
\end{equation}
so that
\begin{equation}
\int_0^\infty \frac{dN_i}{dk}\,dk = N_i.
\end{equation}

For the present benchmark the temperatures satisfy $T_i\gg m_i$, so the spectral maximum occurs close to the relativistic estimate
\begin{equation}
k_{\rm peak}\simeq 1.6\,T_i.
\end{equation}
Accordingly, smaller values of $\delta$ produce larger energy densities, larger temperatures, and spectra whose peaks are shifted toward higher momenta with larger overall normalization.

\begin{figure}[t]
    \centering
    \includegraphics[width=0.48\textwidth]{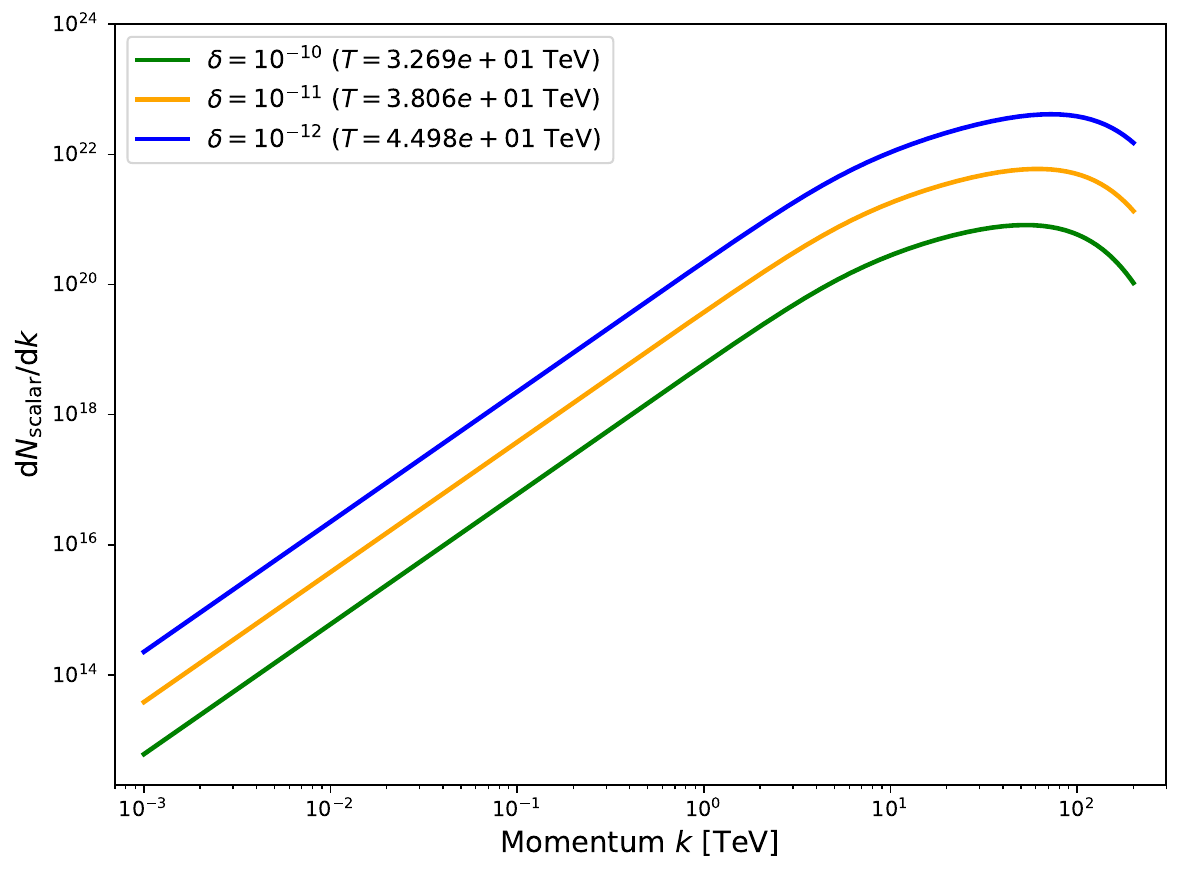}
    \hfill
    \includegraphics[width=0.48\textwidth]{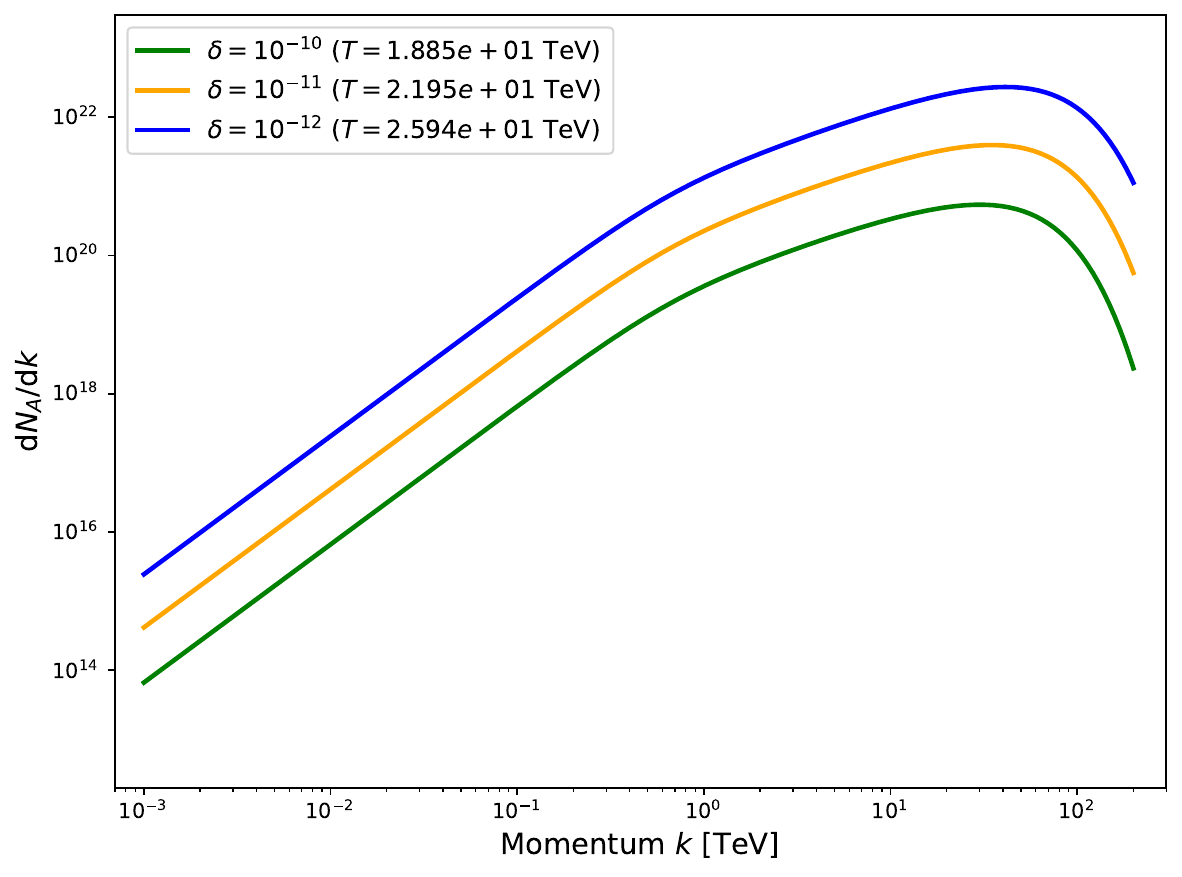}
    \caption{
        Physically normalized thermal spectra $dN_i/dk$ in the broken $U(1)_{\rm EM}$ phase for the scalar and the massive photon. Each panel shows the spectra for $\delta=10^{-12},\,10^{-11},\,10^{-10}$, with larger yields corresponding to smaller $\delta$. The massive-photon spectrum includes only the two transverse polarizations.
    }
    \label{fig:u1em-spectra}
\end{figure}

Figure~\ref{fig:u1em-spectra} shows the fully normalized thermal spectra for the scalar and the massive photon for the three benchmark values of $\delta$. Each curve combines the massive Bose--Einstein shape, the temperature-dependent shift of the peak, and the correct total multiplicity extracted from the microscopic evolution summarized in Table~\ref{tab:u1em-thermal-yields}. The exponential falloff at $k\gg T$ is the usual Boltzmann suppression of the massive tail.

The thermal channel dominates over the vacuum-mismatch contribution by many orders of magnitude. This is physically expected: once the wall becomes ultra-relativistic, even modest friction transfers a substantial fraction of the released vacuum energy into the surrounding medium, and rapid thermalisation converts this energy into an enormous population of relativistic heavy quanta.

In the present $U(1)_{\rm EM}$ setup, the excitations produced in the shocked layer are the heavy scalar and the massive photon. These unstable quanta subsequently decay into Standard Model particles, which then initiate cascades yielding energetic photons, neutrinos, and charged leptons. Thus, once friction is included, the dominant long-range observational signal is expected to arise not from the direct vacuum-mismatch source alone, but from the much larger population of heavy states thermally produced behind the expanding wall.

\section{Phenomenology and observable decay signatures of the broken-phase states}
In order to connect the broken-phase particle content to observable signatures, we first examine the phenomenology of the heavy states that appear once \(U(1)_{\rm EM}\) is broken inside the true-vacuum bubble. In this phase, the relevant excitations are the neutral radial scalar associated with the symmetry-breaking field and the massive photon generated through the Higgs mechanism. Since these unstable states are the primary sources of secondary radiation near the bubble wall, understanding their decays is essential for determining the final signal.

In this section, we therefore study both the neutral radial scalar and the massive photon. We discuss their masses in the broken phase, the effective interactions that allow them to decay, and the dominant channels relevant for our benchmark setup. We then use these decay modes as input to \texttt{Pythia 8} \cite{Sjostrand:2014zea, Bierlich:2022pfr} in order to simulate the subsequent showering, hadronization, and secondary decays, from which we extract the final photon and neutrino spectra.

Because the long-range observables in this scenario are photons and neutrinos rather than the heavy states themselves, our main goal here is to translate the decay properties of the neutral radial scalar and the massive photon into spectra that could in principle be detected far from the nucleation site. As an illustrative example, we show the photon and neutrino number-density spectra obtained from the thermal dissipation channel for the benchmark case \(\delta=10^{-12}\). We focus on this case because thermal production yields far more particles than the vacuum-mismatch mechanism and therefore provides the dominant contribution to the observable signal.

\subsection{Phenomenology of the neutral radial scalar}
\label{cs}

We now consider the phenomenology of the physical scalar excitation associated with the breaking of $\mathrm{U}(1)_{\rm EM}.$ In the present Abelian model, the order parameter is a single complex scalar field $\Phi_{EM}.$ Once $\Phi_{EM}$ acquires a vacuum expectation value, the phase mode is eaten by the photon, which thereby becomes massive, while the only remaining physical scalar degree of freedom is the neutral radial mode. Symmetry breaking in the present $\mathrm{U}(1)_{\rm EM}$ setup therefore leaves only one physical neutral scalar after the Goldstone mode is absorbed.

In the false vacuum (our present universe), the scalar mass is set by the quadratic term in the potential,
\begin{equation}
M_{\rm false}=\sqrt{m^2}=2.58\times10^3~\mathrm{GeV}.
\end{equation}
Inside the true-vacuum bubble, once the scalar develops a nonzero vacuum expectation value \(v\simeq 2191~\mathrm{GeV}\), the curvature of the potential at the minimum gives
\begin{equation}
m_\Phi^2
=
4\lambda v^2+\frac{12\delta v^4}{\Lambda^2}
\qquad\Longrightarrow\qquad
m_\Phi \simeq 5.94\times10^3~\mathrm{GeV}.
\end{equation}

Thus the scalar is relatively light (\(\sim 2.6\) TeV) in the symmetric phase, but becomes substantially heavier (\(\sim 5.9\) TeV) in the broken phase. This large mass splitting reflects the strong dependence of the scalar mass on the vacuum structure.

\paragraph{Dominant decay channels}

Since the physical excitation in the broken phase is a neutral radial scalar, its leading couplings to Standard Model matter are naturally Higgs-like. In the absence of additional light states charged under the broken \(\mathrm{U}(1)_{\rm EM}\), the dominant renormalizable couplings arise through mixing with the Higgs sector, yielding decays into heavy fermions and massive electroweak gauge bosons. Because the scalar is neutral and very heavy, the dominant channels are expected to be those involving the heaviest available Standard Model final states, namely
\begin{equation}
\Phi\to t\bar t,\qquad
\Phi\to W^+W^-,\qquad
\Phi\to ZZ.
\end{equation}

Adopting a Higgs-like effective description, the couplings of the radial scalar to Standard Model fields are rescaled by factors \(\kappa_t\) and \(\kappa_V\) relative to the corresponding Standard Model Higgs couplings at the same mass. For the benchmark adopted here, we take
\begin{equation}
\kappa_t=\kappa_V=0.1.
\end{equation}
The partial width into top quarks is then
\begin{equation}
\Gamma(\Phi\to t\bar t)
=
3\,\frac{\kappa_t^2\,m_t^2\,m_\Phi}{8\pi v_{\rm EW}^2}
\left(1-\frac{4m_t^2}{m_\Phi^2}\right)^{3/2},
\label{eq:width_tt}
\end{equation}
while the electroweak gauge-boson widths are
\begin{equation}
\Gamma(\Phi\to W^+W^-)
=
\kappa_V^2\,
\frac{m_\Phi^3}{16\pi v_{\rm EW}^2}
\sqrt{1-\frac{4m_W^2}{m_\Phi^2}}
\left(1-\frac{4m_W^2}{m_\Phi^2}+12\frac{m_W^4}{m_\Phi^4}\right),
\label{eq:width_WW}
\end{equation}
and
\begin{equation}
\Gamma(\Phi\to ZZ)
=
\kappa_V^2\,
\frac{m_\Phi^3}{32\pi v_{\rm EW}^2}
\sqrt{1-\frac{4m_Z^2}{m_\Phi^2}}
\left(1-\frac{4m_Z^2}{m_\Phi^2}+12\frac{m_Z^4}{m_\Phi^4}\right).
\label{eq:width_ZZ}
\end{equation}

Using
\begin{equation}
m_\Phi=5.94~\mathrm{TeV},
\qquad
m_t=173~\mathrm{GeV},
\qquad
m_W=80.4~\mathrm{GeV},
\qquad
m_Z=91.2~\mathrm{GeV},
\qquad
v_{\rm EW}=246~\mathrm{GeV},
\end{equation}
we obtain
\begin{equation}
\Gamma(\Phi\to t\bar t)\simeq 3.49~\mathrm{GeV},
\end{equation}
\begin{equation}
\Gamma(\Phi\to W^+W^-)\simeq 6.88\times10^2~\mathrm{GeV},
\end{equation}
and
\begin{equation}
\Gamma(\Phi\to ZZ)\simeq 3.44\times10^2~\mathrm{GeV}.
\end{equation}
The total width within this benchmark channel set is therefore
\begin{equation}
\Gamma_{\rm tot}
\simeq 1.04\times10^3~\mathrm{GeV},
\end{equation}
corresponding to
\begin{equation}
\frac{\Gamma_{\rm tot}}{m_\Phi}\simeq 0.17,
\end{equation}
which is consistent with a broad but still well-defined heavy scalar resonance.

The resulting branching ratios within the set of channels included here are approximately
\begin{equation}
{\rm BR}(t\bar t)\simeq 0.34\%,
\qquad
{\rm BR}(W^+W^-)\simeq 66.45\%,
\qquad
{\rm BR}(ZZ)\simeq 33.21\%.
\end{equation}

\paragraph{Omission of the \texorpdfstring{\(\Phi\to hh\)}{Phi to hh} channel}

In addition to the \(t\bar t\), \(W^+W^-\), and \(ZZ\) channels considered here, the decay
\begin{equation}
\Phi\to hh
\end{equation}
is also generically allowed for a heavy neutral scalar. However, its width depends on the cubic scalar coupling \(\lambda_{\Phi hh}\), which is not fixed by the minimal benchmark adopted in this work. In contrast, the \(t\bar t\), \(W^+W^-\), and \(ZZ\) channels can be parameterized directly in terms of the effective Higgs-like couplings \(\kappa_t\) and \(\kappa_V\). We therefore restrict our numerical estimates to the fermionic and electroweak gauge-boson channels, and the quoted branching ratios should be understood within this restricted set of decay modes rather than as the fully inclusive branching fractions of the model.

\begin{table}[h!]
\centering
\begin{tabular}{lcc}
\toprule
\textbf{Decay Mode} & \(\boldsymbol{\Gamma}\) [GeV] & \textbf{BR} [\%] \\
\midrule
\(\Phi\to t\bar t\)   & \(3.49\)              & \(0.34\) \\
\(\Phi\to W^+W^-\)    & \(6.88\times10^{2}\)  & \(66.45\) \\
\(\Phi\to ZZ\)        & \(3.44\times10^{2}\)  & \(33.21\) \\
\midrule
\textbf{Total} & \(\mathbf{1.04\times10^{3}}\) & \textbf{100} \\
\bottomrule
\end{tabular}
\caption{Partial widths and branching ratios for the neutral radial scalar \(\Phi\) at \(m_\Phi=5.94\)~TeV, using the benchmark Higgs-like couplings \(\kappa_t=\kappa_V=0.1\). The quoted branching ratios are normalized only within the restricted channel set \(\Phi\to t\bar t,\ W^+W^-,\ ZZ\). The \(\Phi\to hh\) mode is not included because the cubic coupling \(\lambda_{\Phi hh}\) is not fixed in the minimal benchmark.}
\label{tab:scalar_widths}
\end{table}

Under these assumptions, the decay of the neutral radial scalar is dominated by the electroweak gauge-boson channels, with \(W^+W^-\) providing the largest contribution and \(ZZ\) the next largest one, while the top-quark mode remains subleading. This pattern is the expected one for a very heavy Higgs-like neutral scalar with suppressed but nonzero couplings to the Standard Model.

\subsubsection{Hadronization and final-state yields}

Each neutral-radial-scalar decay produces heavy Standard Model states, dominantly through the channels \(\Phi\to W^+W^-\) and \(\Phi\to ZZ\), with a smaller contribution from \(\Phi\to t\bar t\). These primary decay products then generate hadronic showers and secondary leptons through their subsequent decays. In particular, hadronic decays of the electroweak gauge bosons yield multiple high-$p_T$ jets, while leptonic decays produce charged leptons and neutrinos. When the subleading \(t\bar t\) channel is present, each top quark decays almost exclusively through \(t\to Wb\), thereby adding further \(b\)-initiated jets and additional neutrinos from semileptonic heavy-flavor decays.

As a result of this chain, a single \(\Phi\) decay produces multiple energetic jets, a copious photon spectrum dominated by \(\pi^0\to\gamma\gamma\) from hadron decays, and a significant flux of neutrinos originating from leptonic \(W\) and \(Z\) decays as well as from semileptonic heavy-flavor decays in the hadronic cascade. Because every scalar decay feeds into such a rich electroweak and hadronic final state, the dominant observable signatures are high-energy photons and neutrinos. This makes gamma-ray and neutrino observatories particularly sensitive to scenarios of late-time \(U(1)_{\rm EM}\) breaking.

\begin{figure}[htbp]
  \centering
  \includegraphics[width=0.45\textwidth]{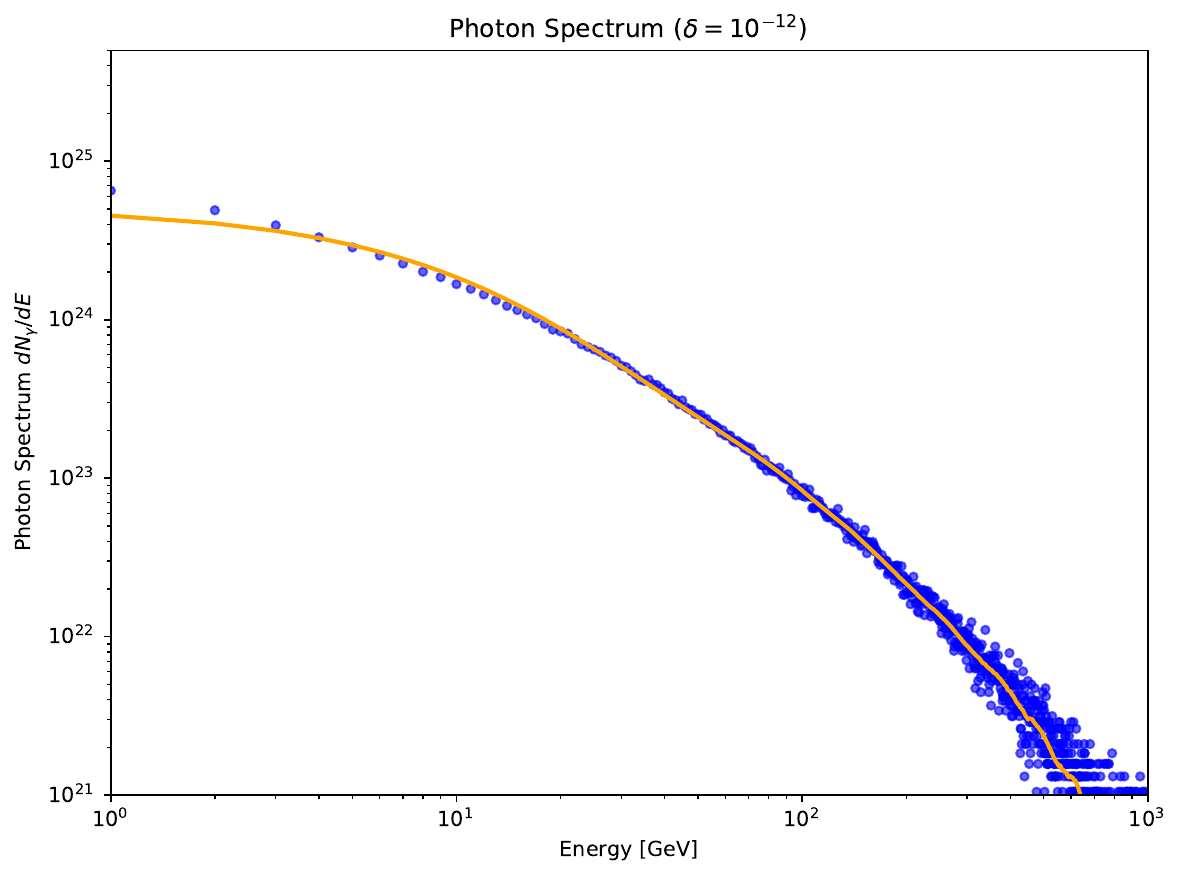}
  \includegraphics[width=0.45\textwidth]{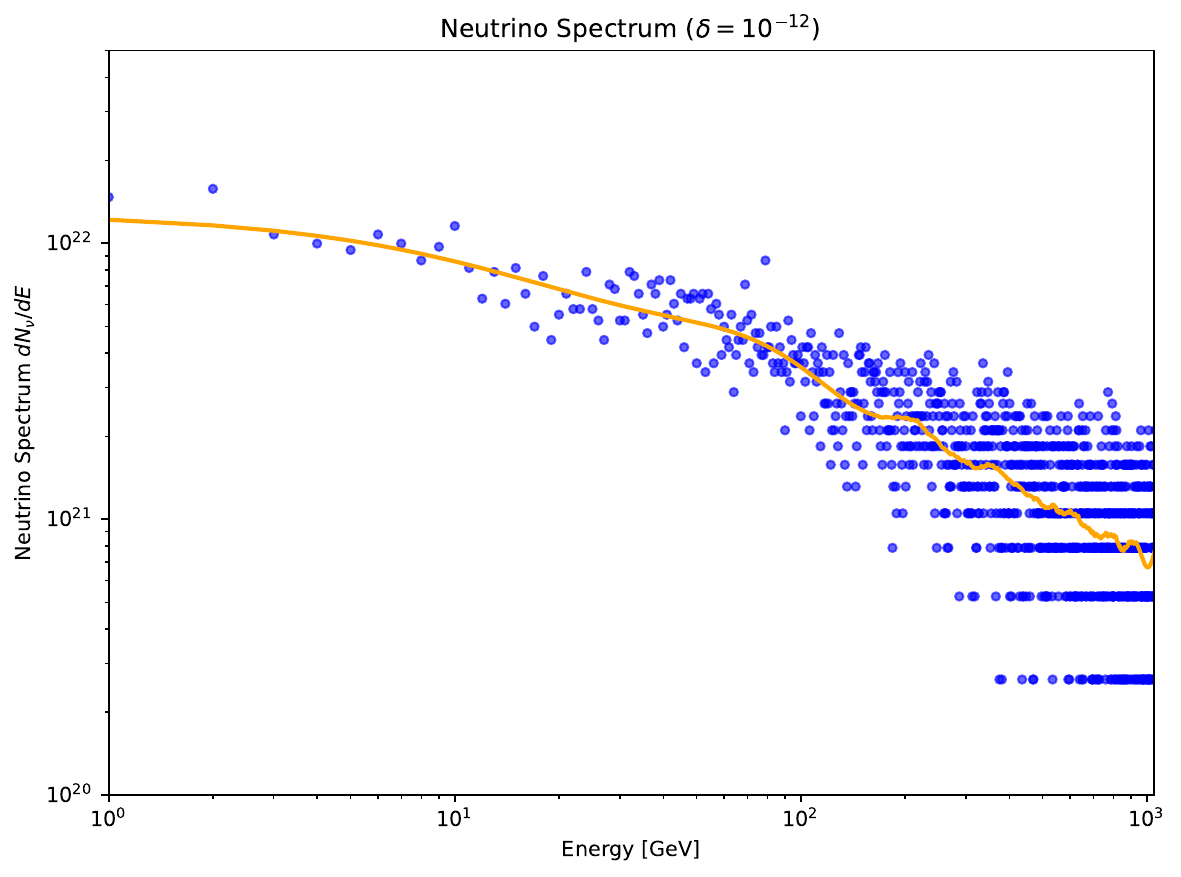}
  \caption{Photon (left) and neutrino (right) energy spectra generated by decays of the \(5.94\) TeV neutral radial scalar produced near true-vacuum bubbles through thermal dissipation. The spectra shown correspond to the benchmark terminal wall velocity characterized by \(\delta = 10^{-12}\), as summarized in Table~\ref{tab:u1em-thermal-yields}.}
  \label{fig:photon_spectrum2}
\end{figure}

Figure~\ref{fig:photon_spectrum2} shows the photon and neutrino energy spectra simulated with \texttt{Pythia} for a \(5.94\) TeV neutral radial scalar.

\subsection{Phenomenology of a massive photon}
When a TeV-scale scalar charged under \(\mathrm{U}(1)_{\rm EM}\) acquires a vacuum expectation value \(v\), the photon becomes massive. In this section we focus on the dominant decay channels of this massive photon, assuming that it retains the usual electromagnetic coupling to charged matter. Possible couplings to electroweak gauge bosons, however, are model-dependent and need not be present at unsuppressed tree level in the minimal setup. We therefore parametrize the \(A W^+W^-\) interaction by an effective coupling \(\kappa\), which should be understood as encoding additional model structure beyond the minimal broken-\(U(1)_{\rm EM}\) framework. For the benchmark adopted here we take \(\kappa=0.3\). This choice provides a representative suppressed coupling for which the \(W^+W^-\) channel remains phenomenologically relevant, while the resulting total width stays below the particle mass so that the resonance description remains meaningful.

At tree level, the partial width into a charged lepton pair is
\begin{equation}
\Gamma(A\to\ell^+\ell^-)
=
\frac{1}{3}\,\alpha(m_A)\,m_A\,
\sqrt{1-\frac{4m_\ell^2}{m_A^2}}\,
\left(1+\frac{2m_\ell^2}{m_A^2}\right),
\label{eq:Gamma_ll_u1}
\end{equation}
where $\alpha(m_A)$ is the running electromagnetic coupling evaluated at the massive-photon scale.

For the inclusive hadronic mode, it is convenient to use the standard high-energy approximation
\begin{equation}
\Gamma(A\to {\rm hadrons})
\simeq
R(m_A^2)\,\Gamma(A\to\mu^+\mu^-)\big|_{m_\mu\to 0},
\label{eq:Gamma_had_u1}
\end{equation}
with
\begin{equation}
R(s)\equiv
\frac{\sigma(e^+e^-\to {\rm hadrons})}{\sigma(e^+e^-\to \mu^+\mu^-)}
\simeq 5
\end{equation}
for $m_A$ well above the light-quark thresholds and above the top threshold. In this approximation,
\begin{equation}
\Gamma(A\to {\rm hadrons})
\simeq
\frac{1}{3}\,\alpha(m_A)\,m_A\,R(m_A^2).
\label{eq:Gamma_had_simple_u1}
\end{equation}

Once $m_A>2m_W$, the two-body decay into electroweak gauge bosons is also open. If an effective trilinear $A W^+W^-$ coupling is present, the corresponding tree-level width (see Appendix~\ref{App-WW}) is
\begin{equation}
\Gamma(A\to W^+W^-)
=
\frac{\kappa^2\alpha\,m_A^5}{192\,m_W^4}\,
\left(1-\frac{4m_W^2}{m_A^2}\right)^{3/2}
\left(
9
-16\frac{m_W^2}{m_A^2}
+48\frac{m_W^4}{m_A^4}
\right).
\label{eq:Gamma_A_WW_explicit}
\end{equation}
For the benchmark adopted here we take \(\kappa=0.3\), so this channel is present but remains parametrically suppressed relative to the unsuppressed EM-like case.

For our benchmark we take
\[
m_A = 657~\mathrm{GeV},
\qquad
m_W = 80.4~\mathrm{GeV},
\qquad
\alpha(m_A)\simeq \frac{1}{128},
\qquad
R(m_A^2)\simeq 5,
\qquad
\kappa = 0.3.
\]
Neglecting the charged-lepton masses to excellent accuracy at this scale, we obtain
\[
\Gamma_e \simeq \Gamma_\mu \simeq \Gamma_\tau
\simeq
\frac{1}{3}\alpha(m_A)m_A
\approx 1.71~\mathrm{GeV},
\]
\[
\Gamma_{\rm had}
\simeq
R\,\frac{1}{3}\alpha(m_A)m_A
\approx 8.56~\mathrm{GeV},
\]
and
\[
\Gamma(A\to W^+W^-)
\approx
\kappa^2 \times 956.15~\mathrm{GeV}
\approx 86.05~\mathrm{GeV}.
\]
The total width is therefore
\[
\Gamma_{\rm tot}
\approx
99.74~\mathrm{GeV},
\]
so that the branching ratios are approximately
\[
{\rm BR}(e^+e^-)\simeq
{\rm BR}(\mu^+\mu^-)\simeq
{\rm BR}(\tau^+\tau^-)\simeq 1.72\%,
\]
\[
{\rm BR}({\rm hadrons})\simeq 8.58\%,
\qquad
{\rm BR}(W^+W^-)\simeq 86.28\%.
\]

\begin{table}[h!]
\centering
\begin{tabular}{lcc}
\toprule
\textbf{Channel} & \(\boldsymbol{\Gamma}\) [GeV] & \(\textbf{BR}\) [\%] \\
\midrule
\(e^+e^-\)           & \(1.71\)   & \(1.72\) \\
\(\mu^+\mu^-\)       & \(1.71\)   & \(1.72\) \\
\(\tau^+\tau^-\)     & \(1.71\)   & \(1.72\) \\
\(\mathrm{hadrons}\) & \(8.56\)   & \(8.58\) \\
\(W^+W^-\)           & \(86.05\)  & \(86.28\) \\
\midrule
\textbf{Total}       & \(\mathbf{99.74}\) & \(\mathbf{100}\) \\
\bottomrule
\end{tabular}
\caption{Partial widths and branching ratios for a \(657\) GeV massive photon, using the running electromagnetic coupling \(\alpha(m_A)\simeq 1/128\), the high-energy ratio \(R(m_A^2)\simeq 5\), and a benchmark effective coupling \(\kappa=0.3\) for the \(A W^+W^-\) interaction.}
\label{tab:widths}
\end{table}

Thus, for the benchmark choice \(\kappa=0.3\), the decay of the massive photon is dominated by the electroweak gauge-boson mode \(A\to W^+W^-\), while the leptonic and hadronic channels remain subleading but phenomenologically relevant for final-state photon and neutrino production.

\subsubsection{Hadronization and final-state yields}

The branching fractions of Table \ref{tab:widths} are fed into \texttt{Pythia} as the hard‐process input. Since the $W^+W^-$ channel dominates, the final photon and neutrino spectra are controlled mainly by hadronic and leptonic W decays, with a smaller contribution from the direct hadronic and leptonic channels.  Neutral pions (\(\pi^0\to\gamma\gamma\)) are produced copiously, while charged pions, kaons, and heavy hadrons yield abundant high-energy neutrinos via semileptonic modes.  Consequently, photons and neutrinos outnumber other stable species by over an order of magnitude, making gamma-ray and neutrino telescopes especially sensitive to TeV-scale massive photon decays.

\begin{figure}[htbp]
  \centering
  \includegraphics[width=0.45\textwidth]{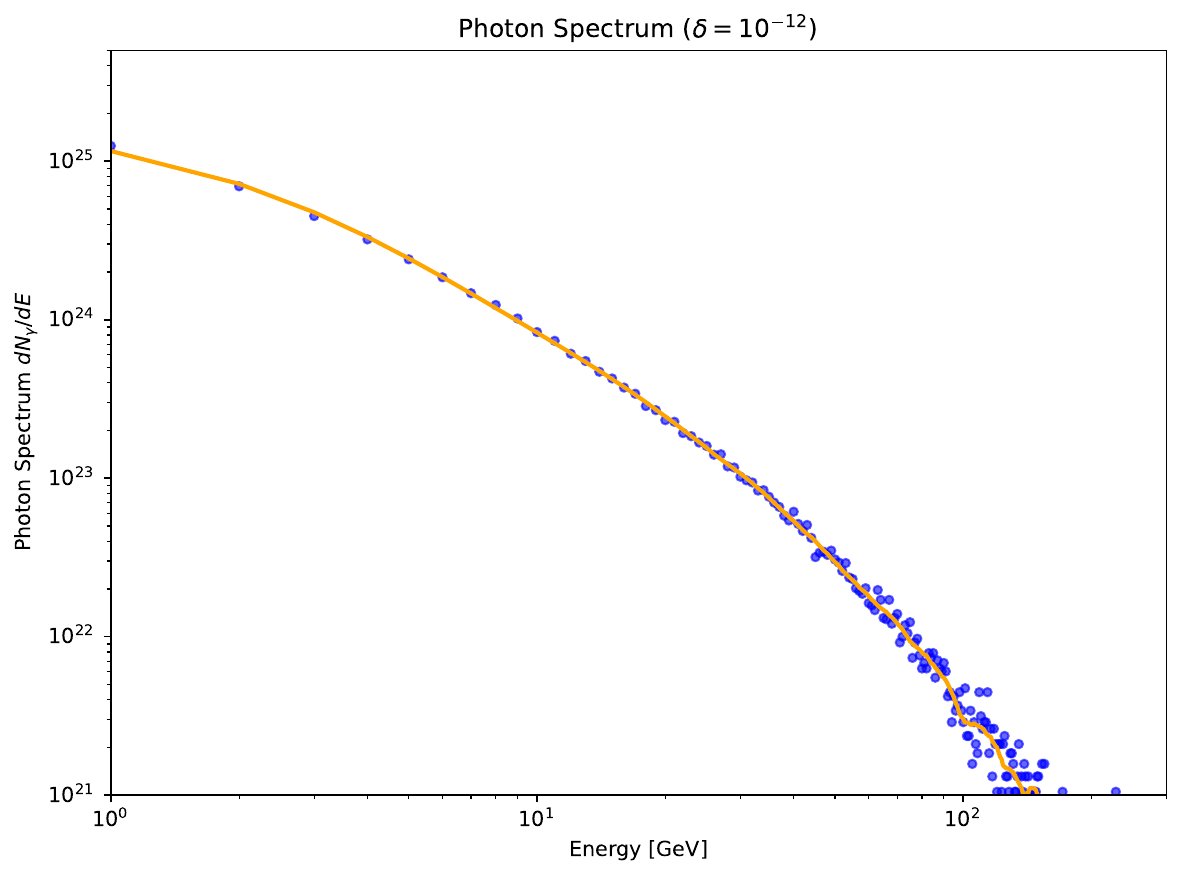}
    \includegraphics[width=0.45\textwidth]{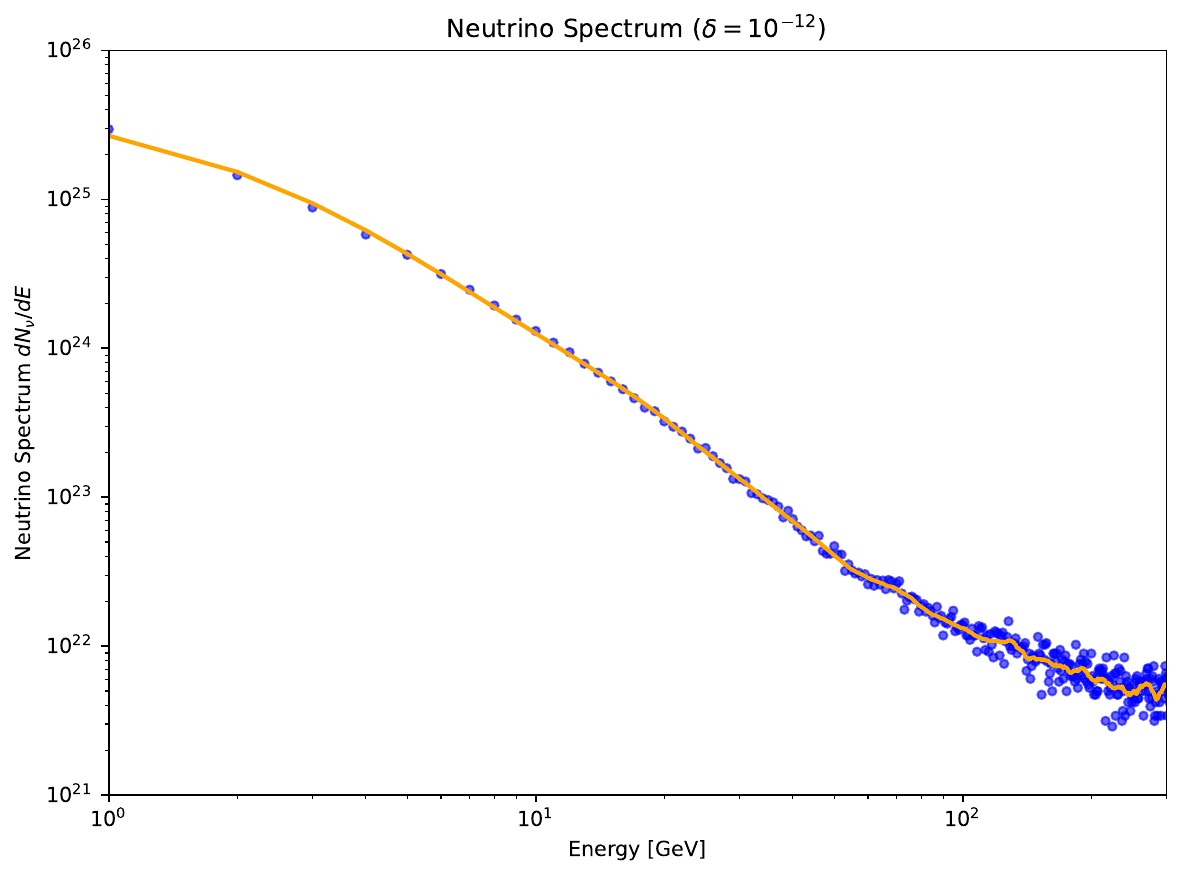}
  {\caption{The spectrum of photons and neutrinos generated by decays of $657$ GeV massive photons produced near true-vacuum bubbles through thermal dissipation. The two panels correspond to the photon spectrum for a terminal wall velocity characterized by $\delta = 10^{-12}$, as summarized in Table~\ref{tab:u1em-thermal-yields}}}
  \label{fig:photon_spectrum3}
\end{figure}

Figures~\ref{fig:photon_spectrum3} display the resulting spectra for photons and neutrinos from a $657$\,GeV photon decay.

The results show that once $U(1)_{\rm EM}$ is broken, the massive-photon channel becomes an important source of excitations in the broken phase. In our conservative thermal estimate, however, we retain only the two transverse polarizations of the massive vector. With this choice, and for the benchmark adopted here, the scalar multiplicity is still somewhat larger because the heating prescription scales with the channel mass parameter $\mu_i.$ Even so, the massive-photon contribution remains phenomenologically very significant and provides a particularly clean potential observational signal alongside the scalar channel.

The spectra shown in Figures~\ref{fig:photon_spectrum2},\ref{fig:photon_spectrum3} for photons and neutrinos represent the number densities at the point of production. To connect these to what would be measured on Earth, one must account for propagation effects. The observable particle flux — defined as the number of particles crossing a unit area per unit time — is reduced by a factor of $1/[4\pi d^{2}(1+z)]$, where $d$ is the physical distance to the source and $z$ its redshift. The extra $(1+z)$ factor arises from relativistic time dilation: two particles emitted with a temporal separation $\Delta t$ at the source are detected with a separation of $(1+z)\Delta t$ at Earth. In addition, the observed particle energies are redshifted and must be corrected by a further factor of $(1+z)$.

\subsection{Phenomenological consistency of the bubble scenario}
\label{subsec:phenomenology-consistency-u1em}

In the $U(1)_{\rm EM}$ scenario, the true vacuum with broken electromagnetism resides inside the bubble, while the familiar unbroken phase persists outside. This configuration naturally determines the associated phenomenology. Within the bubble, the photon acquires a mass $m_A$, and the Higgs mechanism generates longitudinal modes that complete the spectrum of massive vector excitations. As a consequence, the bubble interior supports both massive photons and neutral radial scalar excitations with masses at the symmetry–breaking scale. These unstable broken-phase states cannot propagate unimpeded into the exterior false–vacuum region. Instead, when they interact with the bubble wall, they convert into $U(1)_{\rm EM}$–neutral combinations that subsequently decay into observable photons and neutrinos. Since the bubble interior remains causally hidden until the wall itself arrives, the only long–range signals accessible beforehand are precisely these photons and neutrinos produced on the false–vacuum side of the wall. 

\section{Photon or neutrino signal lead time}
\label{sec:signal_leadtime-u1em}

If bubble walls in this scenario expand at slightly subluminal velocities, then the secondary radiation (photons and neutrinos) can arrive ahead of the advancing wall. These particles therefore constitute a possible ``early warning'' signature of an incoming bubble event. In what follows we quantify this arrival offset for a source located at a distance of one billion light years.

\paragraph{Cosmological framework and arrival delay}

For the benchmark estimate considered here, a full cosmological treatment is not necessary.  
We take a fiducial source at a distance
\begin{equation}
D = 10^9\ {\rm ly},
\end{equation}
which corresponds to a modest redshift, $z\simeq 0.07$, in a flat $\Lambda$CDM cosmology. At such low redshift, the difference between the exact cosmological result and the flat-space estimate is negligible for our purposes, so we use the Minkowski approximation throughout this subsection. If the bubble wall propagates at speed
\begin{equation}
v=(1-\delta)c,
\qquad \delta\ll 1,
\end{equation}
then the arrival delay relative to photons or neutrinos is
\begin{equation}
\Delta t \simeq D\!\left(\frac{1}{v}-\frac{1}{c}\right)
= \frac{\delta}{1-\delta}\,\frac{D}{c}
\simeq \delta\,\frac{D}{c}.
\label{eq:delay-flat-u1em}
\end{equation}
For $D=10^9$ light years, this gives the lead times listed in Table~\ref{tab:photon_delay_u1em}.

\begin{table}[h!]
\centering
\begin{tabular}{|ccc|}
\hline
\textbf{Velocity Deficit $\delta$} & \textbf{Distance (ly)} & \textbf{Time Delay} \\
\hline
$1.0\times10^{-12}$ & $1.0\times10^9$ & 0 d, 8 h, 28 m, 26.35 s \\
$1.0\times10^{-11}$ & $1.0\times10^9$ & 3 d, 12 h, 44 m, 23.52 s \\
$1.0\times10^{-10}$ & $1.0\times10^9$ & 35 d, 7 h, 23 m, 55.25 s \\
\hline
\end{tabular}
\caption{Photon/neutrino lead times for $U(1)_{\rm EM}$ bubble walls with subluminal deficits $\delta$ at a distance of $10^9$ light years, using the flat-space approximation $\Delta t \simeq \delta D/c$. Even for $\delta=10^{-12}$ the lead time is several hours, while larger deficits extend it to days or weeks.}
\label{tab:photon_delay_u1em}
\end{table}

Thus, in the $U(1)_{\rm EM}$ scenario, even tiny departures from luminal wall propagation can generate observable precursor signals. Depending on $\delta$, photons or neutrinos from the decays of the heavy states produced near the wall could reach us hours to weeks before the bubble itself.

\section{Conclusions}

In this work, we investigated the potential cosmological signatures of the late time $U(1)_{EM}$ gauge symmetry breaking. While the $U(1)_{\rm EM}$ symmetry provides masslessness of the
photon  and defines the fundamental structure of our current universe, there is no fundamental principle guaranteeing its eternal persistence. To explore the observational consequences of such a process, we constructed a phenomenological model including a new massive scalar field responsible for the symmetry breaking. The potential of this field facilitates a first-order phase transition driven by the nucleation and expansion of bubbles of true vacuum within the surrounding false vacuum of our current $U(1)_{\rm EM}$-symmetric universe.

A first order phase is associated with copious particle production. The thermal production mechanism across leads to the abundant production of the new scalar field itself, as well as massive photons. We used event generators to simulate the subsequent decay chains of these primary particles. Since the decay channels may include quarks, the final states of these decays were then hadronized using Pythia to obtain the precise spectra of stable particles that would propagate across cosmological distances. Our key finding is that this phase transition would generate a long-range signature dominated by high-energy photons and neutrinos. Therefore, the detection of a specific, diffuse background of photons and neutrinos, inconsistent with any known astrophysical or cosmological source, could serve as a potential indicator of such a phase transition. Consequently, such an observation might be interpreted not merely as evidence of new physics, but as an empirical signal of a fundamental shift in the laws of nature —-- a cosmological "doomsday" event that alters the very forces governing the universe.

In the absence of friction, the bubble walls traveling with the speed of light would arrive at the same time as the signal coming from them. However, a bubble almost always travels through some medium, for example plasma if formed in the early universe or inside stars, or through the interstellar and intergalactic gas. Most importantly, a bubble of true vacuum is engulfed in a sea of particles that produces itself. Therefore, it is not unreasonable to expect that the bubble wall will reach a terminal velocity slightly below the speed of light. Even a very modest slowdown when extrapolated over the cosmological distances may give us some reasonable warning time before the wall hits us. So, if we ever measure spectra like in Figs. \ref{fig:photon_spectrum2}, and 
 \ref{fig:photon_spectrum3}, 
they might represent signals of doomsday. In addition, we note that even when particle production due to vacuum mismatch stops (when the terminal velocity is reached), particles will be produced thermally since a large amount of energy is dumped to the environment due to friction, which is shown in our work. While vacuum-mismatch production provides a useful precursor source and captures the onset of particle emission from the accelerating wall, our analysis shows that the dominant yield comes from friction-induced thermal dissipation in the shocked layer behind the wall; accordingly, the photon and neutrino spectra displayed in this work are based on the thermal channel for the benchmark case $\delta=10^{-12}.$ A more refined treatment including full hydrodynamic backreaction and detailed plasma microphysics would be an interesting extension of the thermal-radiation calculation presented here.

Throughout the paper we used a fiducial value for the energy scale of the phase transition of the order of $1$\,TeV. However, any other value can be used, as long as we are not obviously violating any observational constraint. While the energy scale might be high, the strength of the phase transition (i.e. the difference between the vacua) must be small so that most of the universe is still in the false vacuum today with just a few bubbles here and there. We thus call such phase transitions - late time phase transitions.      

\begin{acknowledgments}
The authors are grateful to L.C.R. Wijewardhana, Jure Zupan, D.C. Dai and Manuel Szewc for carefully reviewing the manuscript and for their valuable comments and suggestions. AS and DS are partially supported by the U.S. National
Science Foundation, under the Grant No. PHY-2310363. AS is also supported by the Grant No. NSF OAC-2417682.
\end{acknowledgments}

\appendix

\section{Appendix~A: Derivation of $\Gamma(A\to W^+W^-)$ from an effective \texorpdfstring{$A W^+W^-$}{AWW} coupling}
\label{App-WW}

In this appendix we derive the decay width for a massive neutral vector boson \(A_\rho\) of mass \(m_A\) into a \(W^+W^-\) pair, assuming the presence of an \emph{effective} trilinear coupling to the charged \(W\) bosons. We emphasize that for a massive photon arising from a broken Abelian \(\mathrm{U}(1)_{\rm EM}\), such a coupling is not automatic in the minimal setup and should be regarded as model-dependent. We therefore parametrize the interaction strength by a dimensionless coefficient \(\kappa\), so that
\begin{equation}
\mathcal{L}_{\rm int}
\,=\,
-\,i \kappa e\,
\Big[
\big(W_{\mu\nu}^- W^{+\mu} - W_{\mu\nu}^+ W^{-\mu}\big) A^\nu
\,+\,
F_{\mu\nu} W^{+\mu} W^{-\nu}
\Big],
\label{eq:L_int_triple}
\end{equation}
where \(F_{\mu\nu}=\partial_\mu A_\nu-\partial_\nu A_\mu\) and
\(W_{\mu\nu}^{\pm}=\partial_\mu W_\nu^{\pm}-\partial_\nu W_\mu^{\pm}\).
For \(\kappa=1\) this reduces formally to the usual EM-like trilinear structure, while \(\kappa\ll1\) corresponds to a suppressed effective coupling.

The corresponding \(A_\rho(p)\,W^+_\mu(p_+)\,W^-_\nu(p_-)\) vertex (all momenta incoming) is
\begin{equation}
i \kappa e\,\Gamma_{\mu\nu\rho}(p_+,p_-,p)\,,
\qquad
\Gamma_{\mu\nu\rho}
=
g_{\mu\nu}(p_+-p_-)_{\rho}
+g_{\nu\rho}(p-p_-)_{\mu}
+g_{\rho\mu}(p_- - p)_{\nu}.
\label{eq:vertex}
\end{equation}

For the decay \(A(k)\to W^+(p_+)\,W^-(p_-)\) we take \(k=p_++p_-\) with \(k^2=m_A^2\), \(p_\pm^2=m_W^2\), and define
\begin{equation}
x \equiv \frac{m_W^2}{m_A^2},
\qquad
\beta \equiv \sqrt{1-4x} \,=\, \frac{|\vec{p}\,|}{E_W}\,,
\qquad
|\vec{p}\,|=\frac{m_A}{2}\sqrt{1-4x},
\qquad
E_W=\frac{m_A}{2}.
\label{eq:x_beta}
\end{equation}

We treat the initial \(A\) as unpolarized and use the Proca/projector polarization sums:
\begin{align}
\sum_{\lambda_A}\varepsilon^{(\lambda_A)}_\rho(k)\,\varepsilon^{*(\lambda_A)}_{\rho'}(k)
&= -\,g_{\rho\rho'} + \frac{k_\rho k_{\rho'}}{m_A^2},\\[3pt]
\sum_{\lambda_+}\varepsilon^{(\lambda_+)}_\mu(p_+)\,\varepsilon^{*(\lambda_+)}_{\mu'}(p_+)
&= -\,g_{\mu\mu'} + \frac{p_{+\mu} p_{+\mu'}}{m_W^2},\\[3pt]
\sum_{\lambda_-}\varepsilon^{(\lambda_-)}_\nu(p_-)\,\varepsilon^{*(\lambda_-)}_{\nu'}(p_-)
&= -\,g_{\nu\nu'} + \frac{p_{-\nu} p_{-\nu'}}{m_W^2}.
\label{eq:proj}
\end{align}

\subsection*{Amplitude and spin sums}

The decay amplitude is
\begin{equation}
\mathcal{M}
=
i \kappa e\,
\varepsilon_A^{\rho}(k)\,
\varepsilon^{*\mu}_{+}(p_+)\,
\varepsilon^{*\nu}_{-}(p_-)\,
\Gamma_{\mu\nu\rho}(p_+,p_-,k).
\label{eq:amp}
\end{equation}
Summing over final polarizations and averaging over the \(3\) initial spin states of \(A\) gives
\begin{equation}
\overline{\sum}|\mathcal{M}|^2
=
\frac{\kappa^2 e^2}{3}\,
\Big( -g_{\rho\rho'}+\frac{k_\rho k_{\rho'}}{m_A^2}\Big)\,
\Big( -g_{\mu\mu'}+\frac{p_{+\mu}p_{+\mu'}}{m_W^2}\Big)\,
\Big( -g_{\nu\nu'}+\frac{p_{-\nu}p_{-\nu'}}{m_W^2}\Big)\,
\Gamma^{\mu\nu\rho}\,\Gamma^{\mu'\nu'\rho'}.
\label{eq:Msq_master}
\end{equation}

\subsection*{Contraction of the polarization projectors}

Using the completeness relations in Eq.~\eqref{eq:proj}, the spin-summed and
spin-averaged squared amplitude becomes
\begin{equation}
\overline{\sum}|\mathcal{M}|^2
=
\frac{\kappa^2 e^2}{3}\,
\Pi^{(A)}_{\rho\rho'}(k)\,
\Pi^{(+)}_{\mu\mu'}(p_+)\,
\Pi^{(-)}_{\nu\nu'}(p_-)\,
\Gamma^{\mu\nu\rho}\Gamma^{\mu'\nu'\rho'},
\end{equation}
where
\begin{equation}
\Pi^{(A)}_{\rho\rho'}(k)= -\,g_{\rho\rho'}+\frac{k_\rho k_{\rho'}}{m_A^2},
\qquad
\Pi^{(\pm)}_{\alpha\beta}(p_\pm)= -\,g_{\alpha\beta}+\frac{p_{\pm\alpha}p_{\pm\beta}}{m_W^2}.
\end{equation}
Carrying out the tensor contraction and expressing the result in terms of
\begin{equation}
x=\frac{m_W^2}{m_A^2},
\qquad
\beta=\sqrt{1-4x},
\end{equation}
one finds
\begin{equation}
\overline{\sum}|\mathcal{M}|^2
=
\frac{\kappa^2 e^2\,m_A^2}{48\,x^2}\,
(1-4x)\,
\bigl(9-16x+48x^2\bigr).
\label{eq:Msq_final_AWW}
\end{equation}

\subsection*{Decay rate}

For a two-body decay into two distinguishable final-state particles,
the standard phase-space formula gives
\begin{equation}
\Gamma(A\to W^+W^-)
=
\frac{|\vec p\,|}{8\pi m_A^2}\,
\overline{\sum}|\mathcal{M}|^2,
\end{equation}
with
\begin{equation}
|\vec p\,|=\frac{m_A}{2}\beta.
\end{equation}
Equivalently,
\begin{equation}
\Gamma(A\to W^+W^-)
=
\frac{\beta}{16\pi m_A}\,
\overline{\sum}|\mathcal{M}|^2.
\end{equation}
Substituting Eq.~\eqref{eq:Msq_final_AWW}, we obtain
\begin{equation}
\Gamma(A\to W^+W^-)
=
\frac{\kappa^2 e^2\,m_A}{768\pi\,x^2}\,
(1-4x)^{3/2}\,
\bigl(9-16x+48x^2\bigr).
\end{equation}
Using \(e^2=4\pi\alpha\), this may be written in the compact form
\begin{equation}
\Gamma(A\to W^+W^-)
=
\frac{\kappa^2\alpha\,m_A}{192\,x^2}\,
(1-4x)^{3/2}\,
\bigl(9-16x+48x^2\bigr),
\qquad
x=\frac{m_W^2}{m_A^2}.
\label{eq:Gamma_A_WW_final}
\end{equation}
Or, restoring the masses explicitly,
\begin{equation}
\Gamma(A\to W^+W^-)
=
\frac{\kappa^2\alpha\,m_A^5}{192\,m_W^4}\,
\left(1-\frac{4m_W^2}{m_A^2}\right)^{3/2}
\left(
9
-16\frac{m_W^2}{m_A^2}
+48\frac{m_W^4}{m_A^4}
\right).
\label{eq:Gamma_A_WW_explicit}
\end{equation}
The decay is kinematically allowed only for
\begin{equation}
m_A>2m_W.
\end{equation}
Near threshold, the width behaves as
\begin{equation}
\Gamma(A\to W^+W^-)\propto \beta^3,
\end{equation}
as expected for a vector decaying into two massive vector bosons. In the
heavy-mass limit \(m_A\gg m_W\), the width scales as
\begin{equation}
\Gamma(A\to W^+W^-)\sim \kappa^2\,\frac{3\alpha}{64}\,\frac{m_A^5}{m_W^4},
\end{equation}
reflecting the enhancement from longitudinal \(W\) polarizations.

The derivation above is valid for the effective interaction in Eq.~\eqref{eq:L_int_triple}. However, for the benchmark mass range considered in the main text, taking \(\kappa\sim1\) can lead to a width comparable to or even larger than \(m_A\), indicating that an unsuppressed EM-like \(A W^+W^-\) coupling is not a self-consistent standalone phenomenological assumption in this regime. In the absence of a UV-complete embedding that controls the longitudinal \(W\) enhancement, this channel should therefore be treated as model-dependent and, if included, parameterized by a suitably suppressed effective coupling \(\kappa\).

\section{\textbf{Appendix~B: Justification of the cutoffs for vacuum--mismatch and thermal production in the $U(1)_{\rm EM}$ transition}}
\label{App-B}

In this appendix we justify the integration cutoffs adopted for the vacuum--mismatch and thermal channels in the $U(1)_{\rm EM}$ benchmark of Secs.~\ref{sec:u1em-prod-spherical} and~\ref{sec:thermal-u1em}. As in the main text, the vacuum--mismatch contribution is terminated once the wall enters the terminal regime, while the thermal sector is evolved up to \(\tau_{\rm final}=5\tau_{\rm term}\). The reason is that the former is controlled directly by the proper acceleration \(\alpha(\tau)\), whereas the latter is governed by the accumulated heating of the shocked shell and therefore does not switch off simply because the acceleration becomes small.

\subsection*{Vacuum--mismatch production: exponential suppression}

The vacuum--mismatch channel is governed by the zero--mode occupation number
\begin{equation}
N_{k=0}(\tau)
=
\left[
\frac{(\omega_+ + \omega_-)^2}{(\omega_+ - \omega_-)^2}
\exp\!\Bigl(\frac{4\omega_+}{\alpha(\tau)}\Bigr)-1
\right]^{-1},
\end{equation}
which, for small positive \(\alpha(\tau)\), reduces to
\begin{equation}
N_{k=0}(\tau)
\simeq
\left[
\frac{(\omega_+ + \omega_-)^2}{(\omega_+ - \omega_-)^2}
\right]^{-1}
\exp\!\left(-\frac{4\omega_+}{\alpha(\tau)}\right).
\end{equation}
Thus the source becomes exponentially small once the acceleration falls below the relevant particle mass scale.

For the $U(1)_{\rm EM}$ benchmark we have
\begin{equation}
\Delta V = 3.460~{\rm TeV}^4,\qquad
\sigma = 3.389~{\rm TeV}^3,\qquad
R_0 = 2.939~{\rm TeV}^{-1},
\end{equation}
so that
\begin{equation}
A\equiv \frac{\Delta V}{\sigma}\simeq 1.021~{\rm TeV},
\qquad
\alpha(0)=A-\frac{2}{R_0}\simeq 0.340~{\rm TeV},
\end{equation}
which is again close to \(1/R_0\). In the late-time regime we may approximate
\begin{equation}
\alpha(\tau)\simeq \alpha(0)\,e^{-\tau/\tau_{\rm term}},
\end{equation}
and therefore at \(\tau=5\tau_{\rm term}\),
\begin{equation}
\alpha(5\tau_{\rm term})=\alpha(0)e^{-5}\simeq 2.29\times10^{-3}~{\rm TeV}.
\end{equation}

For the scalar channel, with \(\omega_+=\mu_s=5.942~{\rm TeV}\),
\begin{equation}
\frac{4\mu_s}{\alpha(5\tau_{\rm term})}\simeq 1.04\times10^4,
\qquad
N_{k=0}^{(s)}(5\tau_{\rm term})\sim e^{-1.04\times10^4},
\end{equation}
while for the massive-photon channel, with \(\omega_+=\mu_\gamma=0.657~{\rm TeV}\),
\begin{equation}
\frac{4\mu_\gamma}{\alpha(5\tau_{\rm term})}\simeq 1.15\times10^3,
\qquad
N_{k=0}^{(\gamma)}(5\tau_{\rm term})\sim e^{-1.15\times10^3}.
\end{equation}
Hence by \(5\tau_{\rm term}\) the vacuum--mismatch source is effectively extinguished in both sectors.

The remaining late-time contribution,
\begin{equation}
\Delta N_{k=0}^{\rm(int)}(5\tau_{\rm term})
=
\int_{\tau_{\rm term}}^{5\tau_{\rm term}}
d\tau\,N_{k=0}(\tau)\,4\pi R^2(\tau)\sinh y(\tau),
\end{equation}
is therefore negligible: although the geometric factor \(R^2(\tau)\sinh y(\tau)\) continues to grow polynomially, this growth is completely overwhelmed by the exponential suppression of \(N_{k=0}(\tau)\). This justifies terminating the vacuum--mismatch evolution once the wall has reached the terminal regime.

\subsection*{Thermal production: sensitivity to the integration time}

The thermal channel behaves differently. Its production rate scales schematically as
\begin{equation}
\frac{dN_i}{d\tau}
\propto
4\pi R^2(\tau)\,\sinh y(\tau)\,n_i[T_i(\tau)],
\qquad
n_i(T)\propto T^3,
\qquad
i=s,\gamma.
\end{equation}
Unlike vacuum--mismatch production, this contribution does not require a large proper acceleration. As long as the shocked shell remains hot and the wall continues to sweep out volume, thermal particle production persists.

In the full simulation the friction coefficient evolves as
\begin{equation}
\eta(\tau)=g_{\rm eff}^2\,T_{\rm eff}^4(\tau),
\end{equation}
with \(\eta(0)\) fixed by the terminal condition. Thus
\begin{equation}
\tau_{\rm term}=\frac{\sigma}{\eta(0)}
\end{equation}
should be interpreted as a reference timescale inherited from the constant-\(\eta\) limit, rather than as an exact stopping time of the nonlinear system. In that simplified limit the accumulated energy deficit approaches its asymptotic value roughly as \(1-e^{-\tau/\tau_{\rm term}}\), so evolving to
\begin{equation}
\tau_{\rm final}=5\tau_{\rm term}
\end{equation}
already captures
\begin{equation}
1-e^{-5}\simeq 0.993
\end{equation}
of the asymptotic deficit. In the full dynamic-\(\eta\) evolution the friction typically increases as the shocked layer heats up, so the approach to the quasi-terminal regime is at least as fast.

Numerically, this choice already captures the dominant thermal yield: extending the integration further changes the final multiplicities only at the few-percent level, whereas stopping substantially earlier would visibly underestimate the result. We therefore adopt \(\tau_{\rm final}=5\tau_{\rm term}\) as a conservative and numerically stable cutoff for the thermal sector. Beyond this point, additional effects such as radiation reaction, scattering losses, and more complete hydrodynamic backreaction of the shocked shell are expected to become increasingly important, while the extra thermal yield within the present framework grows only mildly.

\section{Appendix~C: Proper acceleration of the bubble wall}\label{App-C}

In this appendix we show that the proper acceleration of a frictionless bubble wall is constant and set by the inverse nucleation radius.  
The wall trajectory is described by
\begin{equation}
r^2-t^2=R_0^2,
\end{equation}
which is the standard hyperbola of uniformly accelerated motion in Minkowski spacetime.

A convenient parametrization is in terms of the wall proper time $\tau$,
\begin{equation}
t(\tau)=R_0\sinh\!\left(\frac{\tau}{R_0}\right),
\qquad
r(\tau)=R_0\cosh\!\left(\frac{\tau}{R_0}\right).
\end{equation}
It is immediate to verify that this indeed satisfies the wall trajectory,
\begin{equation}
r^2(\tau)-t^2(\tau)
=
R_0^2\cosh^2\!\left(\frac{\tau}{R_0}\right)
-
R_0^2\sinh^2\!\left(\frac{\tau}{R_0}\right)
=
R_0^2.
\end{equation}
The corresponding four-velocity is
\begin{equation}
u^\mu=\frac{dx^\mu}{d\tau}
=
\left(
\frac{dt}{d\tau},
\frac{dr}{d\tau}
\right)
=
\left(
\cosh\!\left(\frac{\tau}{R_0}\right),
\sinh\!\left(\frac{\tau}{R_0}\right)
\right).
\end{equation}
Its norm is
\begin{equation}
u^\mu u_\mu
=
\left(\frac{dt}{d\tau}\right)^2
-
\left(\frac{dr}{d\tau}\right)^2
=
\cosh^2\!\left(\frac{\tau}{R_0}\right)
-
\sinh^2\!\left(\frac{\tau}{R_0}\right)
=
1,
\end{equation}
confirming that $\tau$ is indeed the proper time along the wall worldline.

Differentiating once more gives the four-acceleration,
\begin{equation}
a^\mu=\frac{du^\mu}{d\tau}
=
\left(
\frac{1}{R_0}\sinh\!\left(\frac{\tau}{R_0}\right),
\frac{1}{R_0}\cosh\!\left(\frac{\tau}{R_0}\right)
\right).
\end{equation}
Its Lorentz-invariant norm is
\begin{equation}
a^\mu a_\mu
=
\left(\frac{1}{R_0}\sinh\!\left(\frac{\tau}{R_0}\right)\right)^2
-
\left(\frac{1}{R_0}\cosh\!\left(\frac{\tau}{R_0}\right)\right)^2
=
-\frac{1}{R_0^2}.
\end{equation}
Therefore the magnitude of the proper acceleration is
\begin{equation}
|a|
=
\sqrt{-\,a^\mu a_\mu}
=
\frac{1}{R_0}.
\end{equation}

Hence, in the absence of friction, the bubble wall undergoes uniformly accelerated motion with constant proper acceleration \cite{Walker:1984vj,Davies:1976hi,Good:2013lca}
\begin{equation}
\alpha=\frac{1}{R_0}.
\end{equation}

\section{Appendix~D: Numerical Procedure}\label{App-D}
To evaluate the friction-limited vacuum-mismatch yield, we solve the coupled system
\(\{y(\tau),t(\tau),R(\tau),N_{\rm tot}(\tau)\}\) as a function of the wall proper time \(\tau\). The evolution is initialized at nucleation with
\begin{equation}
y(0)=0,\qquad t(0)=0,\qquad R(0)=R_0,\qquad N_{\rm tot}(0)=0.
\end{equation}
At each step we compute the instantaneous proper acceleration from
\begin{equation}
\alpha_{\rm raw}(\tau)=A-\frac{2}{R(\tau)}-B\sinh y(\tau),
\end{equation}
where \(A=\Delta V/\sigma\) measures the driving pressure in units of the surface tension, while \(B=\eta/\sigma\) measures the strength of the friction.

As long as \(\alpha_{\rm raw}>0\), the bubble continues to accelerate and we evolve the wall according to
\begin{align}
\frac{dy}{d\tau}&=\alpha_{\rm raw},\\
\frac{dt}{d\tau}&=\cosh y,\\
\frac{dR}{d\tau}&=\sinh y,
\end{align}
together with the particle-production equation
\begin{equation}
\frac{dN_{\rm tot}}{d\tau}
=
N_{k=0}(\tau)\,4\pi R(\tau)^2\sinh y(\tau).
\label{eq:dNtot-u1em-app}
\end{equation}
If \(\alpha_{\rm raw}\le 0\), we interpret this as the onset of terminal motion. In that regime the wall is no longer effectively accelerating, so the vacuum-mismatch source is switched off and the evolution is terminated.

The zero-mode occupation number is evaluated from Eq.~\eqref{eq:Nk0-alpha-u1em}. For sufficiently large values of \(4\omega_+/\alpha_{\rm raw}\), however, the direct exponential form becomes numerically stiff. In that case we replace it by the asymptotic approximation
\begin{equation}
N_{k=0}(\tau)\simeq
\left[\frac{(\omega_++\omega_-)^2}{(\omega_+-\omega_-)^2}\right]^{-1}
e^{-4\omega_+/\alpha_{\rm raw}},
\end{equation}
which captures the same late-time exponential suppression while remaining numerically stable.

The integration is performed with adaptive stepping in \(\tau\). In practice, the step size is chosen from the local acceleration scale, so that smaller steps are used when the evolution is rapid and larger steps are allowed once the system begins to approach terminal balance. We also impose upper and lower bounds on the step size in order to keep the evolution stable throughout both the early curvature-dominated stage and the late near-terminal stage. Convergence was checked by repeating the evolution with smaller steps and confirming that the quantities \(\tau_{\rm term}\), \(R_{\rm fin}\), and \(N_{\rm tot}^{\rm(int)}\) remain stable to better than \(10^{-4}\).

For a given benchmark value of the terminal deficit
\begin{equation}
\delta=1-v_{\rm term},
\end{equation}
we determine the corresponding friction coefficient by imposing the terminal-balance relation
\begin{equation}
\gamma_{\rm term}v_{\rm term}=\frac{\Delta V}{\eta}.
\end{equation}
This gives
\begin{equation}
\eta(\delta)=\frac{\Delta V}{\gamma_{\rm term}v_{\rm term}},
\qquad
\tau_{\rm term}=\frac{\sigma}{\eta}.
\end{equation}
In this way each choice of \(\delta\) fixes a unique friction scale and therefore a unique wall trajectory.

Throughout the numerical evolution we monitor the behaviour of \(R(\tau)\), \(y(\tau)\), and \(N_{k=0}(\tau)\). In all cases considered here the radius grows monotonically, while the occupation number decreases rapidly once the friction term begins to compete with the vacuum drive. The final integrated yield therefore reflects the competition between two effects: the geometric enhancement coming from the factor \(4\pi R^2\sinh y\), and the exponential suppression of \(N_{k=0}\) as the proper acceleration drops. This is precisely the interplay that the numerical calculation is designed to capture.
\bibliographystyle{JHEP}
\bibliography{References}
\end{document}